\documentclass[preprints,article,communication,accept,moreauthors,pdftex,10pt,a4paper]{Definitions/mdpi} 
\firstpage{1} 
\pubvolume{xx}
\issuenum{1}
\articlenumber{5}
\pubyear{2019}
\copyrightyear{2019}
\history{Received: date; Accepted: date; Published: date}

\pdfoutput=1

\usepackage{dcolumn}   
\graphicspath{ {figs/} } 
\usepackage{slashed} 
\usepackage[utf8]{inputenc}
\usepackage{ amssymb } 
\usepackage{xcolor} 
\usepackage{bm} 
\graphicspath{ {figs/} } 

\newcommand{\Lagr}{\mathcal{L}}

\newcommand{\mev}{\text{MeV}}

\Title{Estimating the variation of neutron star observables by symmetric dense nuclear matter properties}

\Author{P\'eter P\'osfay$^{1,\ddagger}$*, Gergely G\'abor Barnaf\"oldi$^{1,\ddagger}$\orcidB and Antal Jakov\'ac$^{2,\ddagger}$\orcidC}

\AuthorNames{P\'eter P\'osfay,  Gergely G\'abor Barnaf\"oldi and Antal Jakov\'ac}

\address{%
$^{1}$ \quad Wigner research Centre for Physics of the Hungarian Academy of Sciences\\
$^{2}$ \quad Eötvös Loránd University, Budapest}

\corres{Correspondence: posfay.peter@winger.mta.hu}

\secondnote{These authors contributed equally to this work.}

\abstract{Recent multi-channel astrophysics observations and the soon-to-be published new
measured electromagnetic and gravitation data provide information on the inner structure of the compact stars. These macroscopic observations can significantly increase our knowledge on the neutron star enteriors, providing constraints on the microscopic physical properties. On the other hand, due to the masquarade problem, there are still uncertainties on the various nuclear-matter models and their parameters as well. 
Calculating the properties of the dense nuclear matter, effective field theories are the most widely-used tools. However the values of the microscopical parameters need to be set consistently to the nuclear and astrophysical measurements.
In this work we investigate how uncertainties are induced by the variation of the microscopical parameters. We use a symmetric nuclear matter in an extended $\sigma$-$\omega$ model. We calculate the dense matter equation of state and give the mass-radius diagram. We present that the Landau mass and compressibility modulus of the nuclear matter have definite linear relation to the maximum mass of a Schwarzschild neutron star.}


\begin{document}


\section{Introduction}

The investigation of the structure of compact astrophysical objects like neutron stars, magnetars, quark- or hybrid stars {\sl etc.} is an active novel research area as a child of astrophysics, gravitational theory and experiment and nuclear physics. So far the extreme dense state of the matter can not be produced in today's Earth-based particle accelerators, thus only celestial objects can be used for tests.
Electromagnetic measurements, such as X-ray- and gamma satellites, aim to measure properties of these objects more and more accurately~\cite{nasanicer,Merloni:2012uf,athena,Ozel:2015ykl}. In parallel, radio array data~\cite{Watts:2019lbs} and the newly discovered gravity waves provide a new way to probe their inner structure~\cite{Abott_gw1708_radius,LIGO_NSNSGW_detection,Rezzolla:2016nxn}. These observations are particularly important inputs for the theoretical studies of dense nuclear matter~\cite{Ozel:2016,Raithel:2017ity}. 

From the theoretical point of view, first principle calculations based on lattice field theory are still challenging at high chemical potentials present in compact stars~\cite{Katz_finite_mu_analitic_cont,katz_finite_mu_EoS,Katz_finite_mu_lattice}. Thus, 
 effective theories play an important role in studying the properties of cold dense nuclear matter~\cite{Holt:2014hma, Kojo:2017pfw}.
Recent studies show the importance of the correct handling of the bosonic sector in effective theories of nuclear matter~\cite{harmonic:2017,Zsolt:2007}, moreover, applying the functional renormalization group (FRG) method on the simplest non-trivial nuclear matter, the effect of the microscopical parameters on neutron star observables were shown in Refs.~\cite{Posfay:2017cor,Posfay:2016ygf}.

In this paper we study the connection between the parametrizations of effective nuclear models and measurable properties of compact stars in three differently extended versions of the $\sigma$-$\omega$ model. All these include symmetric nuclear matter with various interaction terms in the bosonic sector. After calculating the equation of state (EoS) corresponding to different parametrizations of these models, the mass-radius ($M$-$R$) diagrams are calculated by solving the Tollmann\,--\,Oppenheimer\,--\,Volkoff (TOV) equations. We show how sensitive the mass-radius relation is to differences in the bosonic sector. The dependence of particular properties of compact stars (maximum mass and radius) is presented, influenced by different saturation parameters of the symmetric nuclear matter. 

\section{The extended $\sigma$-$\omega $ in the mean field approximation}

Here we apply the most common mean field model of the dense nuclear matter, formulating the extended $\sigma$-$\omega$ model~\cite{WALECKA1974491,PhysRev.98.783} with the Lagrange-function taken from Refs.~\cite{norman1997compact,Schmitt:2010},
%
\begin{equation}
\Lagr= N_{f} \,
\overline{\Psi} \left( i \slashed{\partial} -m_{N} + g_{\sigma} \sigma -g_{\omega} \slashed{\omega} \right) \Psi 
+\frac{1}{2}\,\sigma \left(\partial^{2}-m_{\sigma}^2 \right) \sigma - U_{i}(\sigma) 
- \frac{1}{4}\,\omega_{\mu \nu} \omega^{\mu\nu}+\frac{1}{2}m_{\omega}^2\omega^2,
\label{eq:wal_lag}
\end{equation}
%
where $\Psi$ is the fermionic nucleon field,  $N_{f}=2$ is the number of nucleons, and $m_{N}$, $m_{\sigma}$, and $m_{\omega}$ are the nucleon, sigma, and omega masses, respectively, for the usual scalar and vector fields. We introduced the  $\omega_{\mu \nu}=\partial_{\mu} \omega_{\nu}-\partial_{\nu} \omega_{\mu}$, and the Yukawa coupling corresponding to the $\sigma$-nucleon and $\omega$-nucleon interactions is given by $g_{\sigma}$ and $g_{\omega}$. We denote the general bosonic interaction terms with $U_{i}(\sigma)$ which can have thee different forms as the considered modified model cases for certain $i$,
\begin{equation}
\begin{aligned}[lc]
U_{3} &= \lambda_{3} \sigma^{3}, \\
U_{4} &= \lambda_{4} \sigma^{4}, \\
U_{34}&=\lambda_{3} \sigma^{3} + \lambda_{4} \sigma^{4} .
\end{aligned}
\label{eq:U_types}
\end{equation}
%

In the mean field (MF) approximation the kinetic terms are zero for the mesons and only the fermionic path integral has to be calculated at finite chemical potential and temperature. We consider here the symmetric nuclear matter to be in equilibrium which includes the baryon number conservation. Taking into account this, the standard procedure were applied by minimizing the free energy of the infinite symmetric nuclear matter at the zero temperature limit, where for the proton ($n_p$) and neutron ($n_n$) number densities are equal, such as the proper chemical potentials, $\mu_p$ and $\mu_n$ respectively.
\begin{equation}
n_p = n_n \quad \quad \longrightarrow \quad \quad \mu_p=\mu_n = \mu \ .  
\end{equation}
After applying this for all the three cases in eq.~\eqref{eq:U_types} substituting into eq.~\eqref{eq:wal_lag}, numerical solution can be obtained after parameter fitting.

\section{Parameter fitting in the extended $\sigma$-$\omega$ model}

As the general procedure, all the models considered cases in  eq.~\eqref{eq:U_types} need to fit to the nucleon saturation data found in e.g. Refs.~\cite{norman1997compact,meng2016relativistic}. 
In parallel to the effective mass, we introduced the definition of the Landau mass
%
\begin{equation}
m_{L} =\frac{k_{F}}{v_{F}}  \quad \quad \textrm{with} \quad \quad  v_{F} =\left.\frac{\partial E_{k}}{\partial k} \right|_{k=k_{F}}.
\label{eq:landau_mass}
\end{equation}
%
Where $k=k_{F}$ the Fermi-surface and $E_{k}$ is the dispersion relation of the nucleons. The Landau mass ($m_L$) and the effective mass ($m^{\ast}$) are not independent in relativistic mean field theories, 
%
\begin{equation}
m_{L}= \sqrt{k_{F}^2 + {m^{\ast}}^2} \ .
\label{eq:effmass_vs_landau_mass}
\end{equation}
%
%
This is the reason why the Landau mass and the effective mass of the nucleons can not be fitted simultaneously in the models we consider~\cite{meng2016relativistic}. In this paper we deal with this problem in the following way. We fit all of the models two times: using the effective mass value from Table~\ref{tab:fitting_data} and one calculated from eq.~\eqref{eq:effmass_vs_landau_mass} to reproduce the Landau mass value from Table~\ref{tab:fitting_data}. 
%
%
\begin{table}[H]
\caption{Nuclear saturation parameter data, from Refs.~\cite{norman1997compact,meng2016relativistic}.}
\label{tab:fitting_data}
\centering
\begin{tabular}{lll}
\toprule
\textbf{Parameter}              & \textbf{ Value}  & \textbf{Unit}      \\
\midrule
Binding energy, $B$         & $-16.3$ & MeV    \\
Saturation density, $n_{0}$     & 0.156 & fm\textsuperscript{-3}  \\
Nucleon effective mass, $m^{\ast}$ & 0.6 $m_{N}$  &  MeV      \\
Nucleon Landau mass, $m_L$    & 0.83 $m_{N}$ & MeV \\
Incompressibility, $K$      & 240 & MeV   \\  
\end{tabular}

\end{table}

If the models with $U_{3}$ and $U_{4}$ type interaction terms are used, than there is not enough free parameters to fit the data in Table~ \ref{tab:fitting_data}. In these cases the nucleon effective mass, saturation density, and binding energy are fitted and the compression modulus is a prediction, given by
\begin{equation}
K =k_{F}^2 \frac{\partial^2 }{\partial k_{F}^2}(\varepsilon /  n)  = 9 n^2 \frac{\partial^2 }{\partial n^2}(\varepsilon / n) \ ,
\label{eq:K}
\end{equation}
which has a simple connection to the thermodynamical compressibility at the saturation density $n_{0}$.
%
%

In the case of $U_{34}$, all four parameters can be fitted simultaneously, and there is another way to incorporate data regarding both Landau and effective mass. For this model we consider a third fit, where the value of the effective mass is chosen in a way that minimizes the error coming from not fitting the two types of masses correctly. Technically this value of the effective mass minimized the $\chi^2$ of the fit, with value
%
%
\begin{equation}
m_{opt}= 0.6567 \, m_{N} \approx 616 \, \mev \ .
\label{eq:optimal_mass}
\end{equation}

Since the incompressibility is different for the model cases with different interactions terms, we compared them in Table~\ref{tab:K_comp}. For model cases with $U_{3}$ and $U_{4}$ there are two fits, for Landau and effective mass which produce different incompressibility values, because they do not have enough free parameters to fit the correct value. However, for $U_{34}$ there are enough parameters to fit the incompressibility, so it has the same value for all three fits: for the Landau mass, for the effective mass and for the optimal mass. As Table~\ref{tab:K_comp} presents incompresibility values for $U_{3}$ with Landau fit is quite close and $U_{34}$ result provide the best fit with the saturation nuclear matter parameters in Table~\ref{tab:fitting_data}. These models differ in their predictions for higher densities of nuclear matter which complicates the description of the compact star interior. 
%
\begin{table}[H]
\caption{ The obtained incompressibility values in different model cases and fits.}
\label{tab:K_comp}
\centering
\begin{tabular}{llc}
\toprule
\textbf{Models}    & \textbf{Calculation method}          & \textbf{K [MeV]}        \\
\midrule
$\sigma$-$\omega$ model & reference value & 563    \\
$U_{3}$ & effective mass fit    & 437   \\
$U_{3}$ & Landau mass fit &  247  \\
$U_{4}$ &  effective mass fit  & 482 \\
$U_{4}$ &  Landau mass fit     & 334   \\  
$U_{34}$  & $\chi^2$ fits for all   & 240   \\  
\end{tabular}

\end{table}

\section{Properties of nuclear matter in the extended $\sigma$-$\omega$ model}

The nuclear properties of the different model cases were compared with all the possible parameter fits at the equation of state (EOS) level. We used all three types of interaction terms: $U_{3}$, $U_{4}$, and $U_{34}$, and each was considered with two parametrizations corresponding to Landau and effective mass fits. All results were cross-checked with the original $\sigma$-$\omega$ model parameters. In case of the model characterized by $U_{34}$ interactions, we used a fit which reproduces \eqref{eq:optimal_mass}. 

\begin{figure}[!h]
\begin{center}
\includegraphics[width=0.80\textwidth]{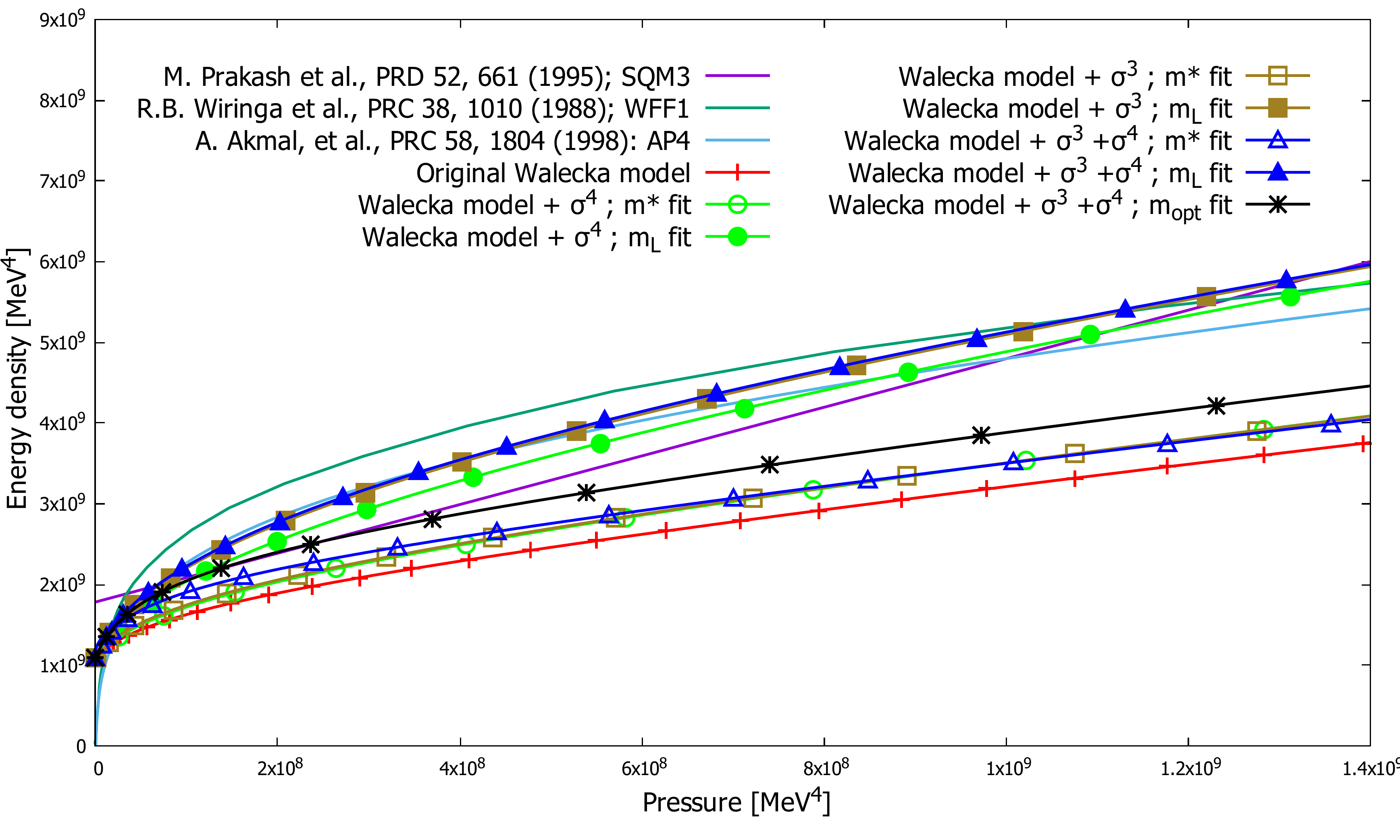}
\end{center}
\vspace*{-0.5 truecm}
\caption{\label{fig:eos_comp}Equation of state in different models in comparision to Refs.~\cite{PhysRevD.52.661,PhysRevC.38.1010,PhysRevC.58.1804}.}
\end{figure}

The energy density, pressure and density were calculated in all of these models. The equation of states corresponding to these model and fit cases are shown on Fig.~\ref{fig:eos_comp}. The results from the modified $\sigma$-$\omega$ model are compared to other equation of state parametrizations from Refs.~\cite{PhysRevD.52.661,PhysRevC.38.1010,PhysRevC.58.1804} ({\sl solid lines}). An important feature of Figure~\ref{fig:eos_comp} is that different model based EoS parametrizations are separated based on whether they are parametrized by the Landau ({\sl full symbols}) or the effective mass ({\sl open symbols}). The models which are fitted to reproduce the correct effective mass of nucleons has smaller energy density at given pressure. This phenomena becomes more prominent as pressure increases. It also important to note that in a given band the incompressibility corresponding to certain equation of states can be very different. For example in the group which fitted for the effective mass in the $U_{3}$ type model ({\sl open rectangles}) has the incompressibility $K=$ 247~MeV, but the $U_{4}$ type ({\sl open circles}) has almost double value, $K=$ 482~MeV.
%
%

\begin{figure}[!h]
\begin{center}
\includegraphics[width=0.80\textwidth]{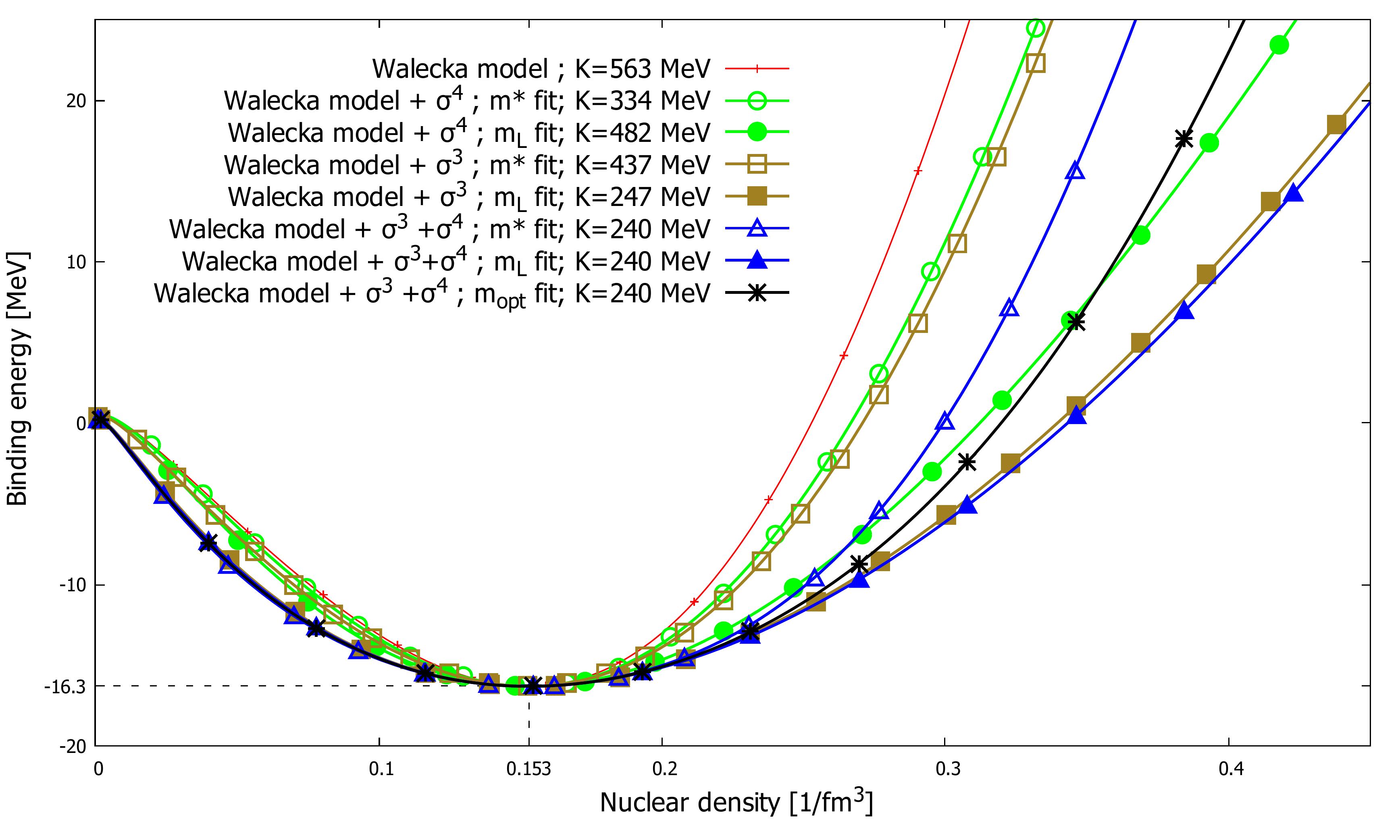}
\end{center}
\vspace*{-0.5 truecm}
\caption{\label{fig:bound_comp} The density dependence of binding energy with the different models and parametrizations.}
\end{figure}

Similar grouping can be seen on Figure~\ref{fig:bound_comp} presenting the binding energy $B$ as the function of the nuclear density $n$. However, near the minimal $B$, the value of incompressibility governs the curves, because it determines the curvature of the curves around minimum. Effective mass fits with {\sl open symbols} has steeper rise with the increasing density, $n$, while Landau-mass fit curves with {\sl full symbols} are wider. We note that below the saturation density curves are getting more independent on the model choice and parameter fits.   
%

\section{Compact star observables from the effective models}

The Tolman\,--\,Oppenheimer\,--\,Volkoff equations provide the general relativistic description of the compact stars assuming spherically symmetric and time independt space-time structure~\cite{norman1997compact}, 
%
%
\begin{equation}
\begin{aligned}
\frac{d P(r)}{d r} & =- \frac{G \varepsilon(r) m(r)}{r^2} 
\left[ 
1 + \frac{P(r)}{\varepsilon(r)}
\right]
\left[
1 + \frac{4 \pi r ^3 P(r)}{m(r)}
\right]
\left[
1 - \frac{2 G m(r)}{r}
\right]^{-1}  \ , \\
\frac{d m(r)}{ d r } & =4 \pi r^2 \varepsilon(r)  \ .
\end{aligned}
\label{eq:TOV}
\end{equation}
%
%
Here $P(r)$  and $\varepsilon(r)$ are the pressure and energy density as functions of the radius of the star, $G$ is the gravitational constant while $m(r)$ is the mass of star which is included in a shell with radius $r$. To integrate the equations, one needs a connection between $P(r)$ and $\varepsilon(r)$ at given $r$ which is provided by the nuclear matter equation of state in the form of the relation $P(r)=P(\varepsilon(r))$. To start the integration one has to choose a central energy density value $\varepsilon_{c}$ for the star as an initial condition.

\begin{figure}[!h]
\begin{center}
\includegraphics[width=0.90\textwidth]{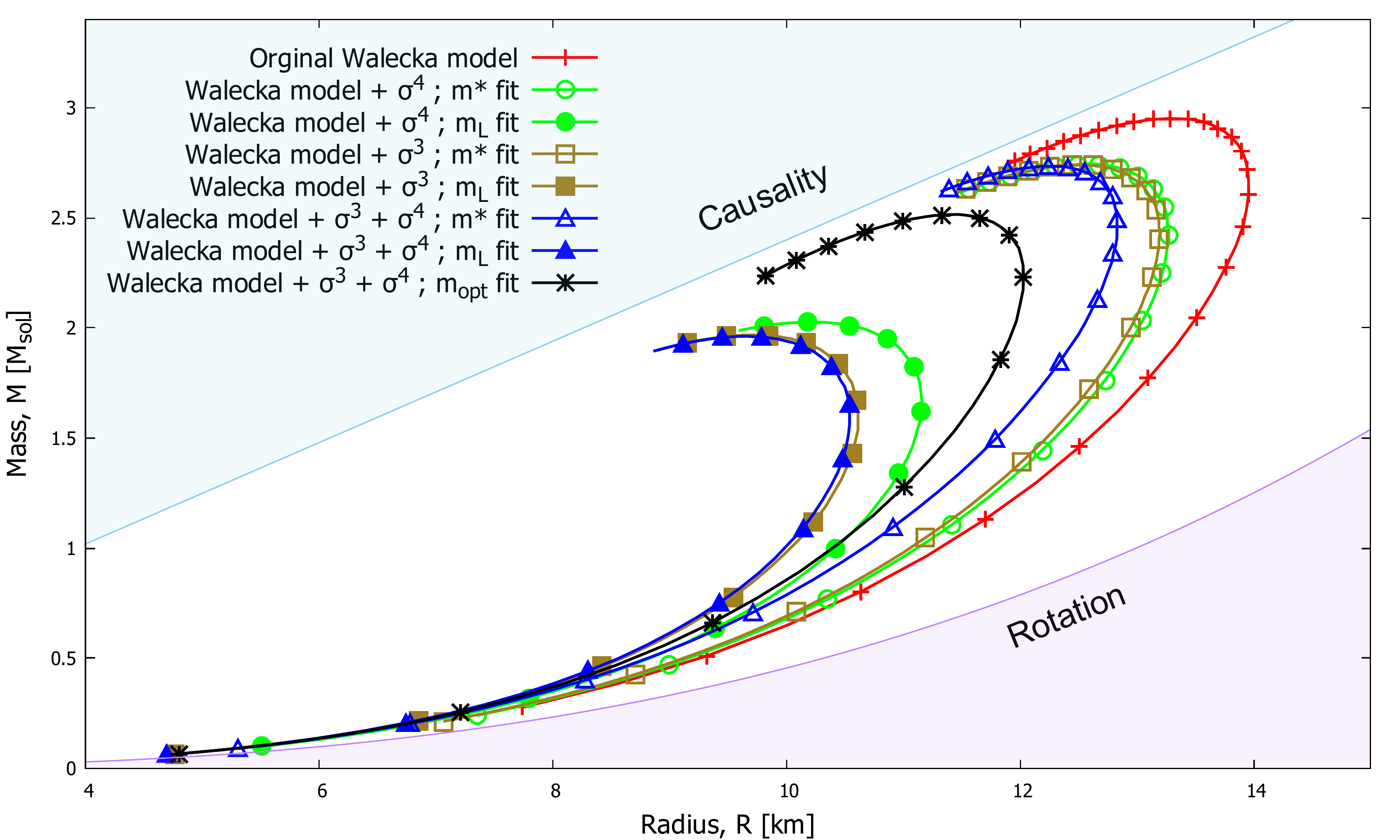}
\end{center}
\vspace*{-0.5 truecm}
\caption{\label{fig:mr} Mass - radius diagram corresponding to the different parametrizations of the modified $\sigma$-$\omega$ model. The line corresponding to the original $\sigma$-$\omega$ model is also drawn for comparison with {\sl red color}.}
\end{figure}

After solving the equations~\eqref{eq:TOV} using the EoS from the model cases with various fits, the mass ($M$) and radius ($R$) of a compact star with a given energy density can be determined. Results corresponding to different energy densities in a given model are summarized on a mass-radius $M$ - $R$ diagram on Figure~\ref{fig:mr}. 
%
%

The model variants inherited the behaviour as on Figs.~\ref{fig:eos_comp} and~\ref{fig:bound_comp}: the curves are grouped based on whether or not they parametrized to reproduce the effective mass or Landau mass. Models with smaller effective mass ({\sl open symbols}) systematically produce higher maximum mass stars compared to their parametrization with larger effective mass (Landau mass, with {\sl full symbols}). Moreover, all models fitted for the effective mass value in Table~\ref{tab:fitting_data} produce higher maximum star mass than the ones fitted for the Landau mass. Since the Landau mass and effective mass are not independent as in eq.~\eqref{eq:landau_mass}, the above statement is equivalent to saying that higher effective mass produces smaller maximum star mass. This picture is supported by the curve corresponding to the model case parametrized by the optimal mass in eq.~\eqref{eq:optimal_mass} which is the best fit of the model. The maximum star mass in this case is between the values produced by parametrizations described by smaller and larger effective nucleon mass. It is interesting to note that parametrizations corresponding to the effective mass and to the optimal mass value may be ruled out by observations, as the produce much larger
maximum star mass and radius than the most recent measurements and
theoretical predictions suggest~\cite{Ozel:2016oaf}.
 
The value of incompressibility in these models seems to influence very little the maximum star mass, however, at the same time it has an effect on the compactness by influencing the radius of the star. Thus can be seen on Fig.~\ref{fig:mr} by looking at the curves belonging to the fits with the same effective mass (Landau mass). In these model cases all of the parameters are the same which listed in Table~\ref{tab:fitting_data} apart form the incompressibility. 

\section{Connection between maximum star mass and nuclear parameters}

\begin{figure}[!h]
\begin{center}
\includegraphics[width=0.81\textwidth]{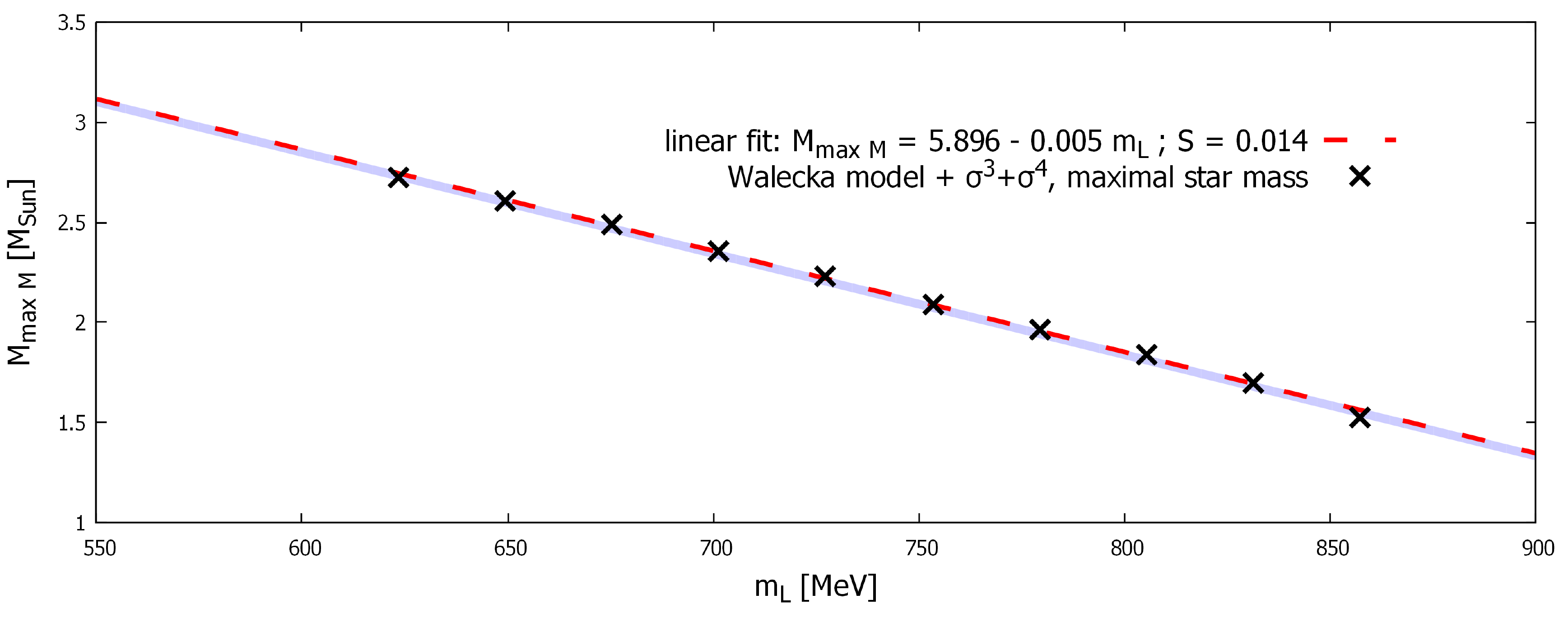}
\includegraphics[width=0.81\textwidth]{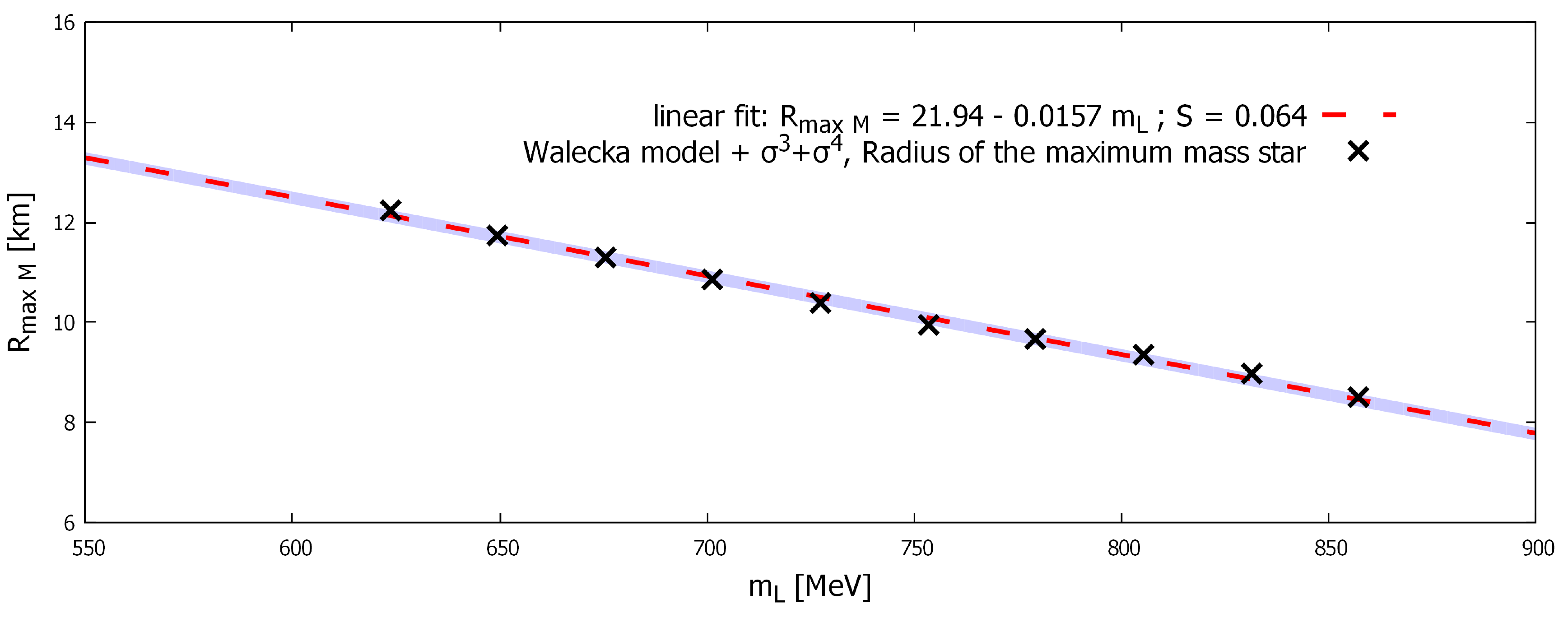}
\end{center}
\vspace*{-0.5 truecm}
\caption{\label{fig:R_ml} Maximum star mass, 
$M_{max\,M}$ ({\sl upper panel}) and maximum star size, $R_{max\,M}$ ({\sl lower panel}) as function of the Landau nucleon mass, $m_L$.}
\end{figure}

Our results present strong connection between the nuclear effective mass, $m_L$ and maximum star mass, $M_{max\,M}$. To further study this phenomena, the model variant with interaction term $U_{34}$ is used and presented below. The couplings in the model are calculated to reproduce all values listed in Table~\ref{tab:fitting_data} except for the nuclear effective mass and Landau mass which are kept as external parameters. To study the maximum star mass dependence on effective nucleon mass, many different fits of this model are considered all with different values of effective nucleon mass. The $M$ - $R$ diagrams corresponding to these parametrizations are calculated, and the maximum star mass is determined in each case. This procedure makes it possible to determine the dependence of maximum star mass on nucleon effective mass in this model, if everything else is kept constant. The results are summarized in Fig.~\ref{fig:R_ml}. The connection between maximum star mass and nucleon effective mass is well described by a linear connection, which gives the best fit of the numerical data,
%
%
\begin{equation}
M_{max\, M}= 5.896 - 0.005 \, m_{L}
\label{eq:M_ml_fit}
\end{equation}
%
%
where $M_{max\,M}$ is given in units of $M_{Sun}$ and $m_{L}$  is given in MeV.

Although the above equation is derived considering only the model with interaction term $U_{34}$ and taking into account parametrizations which differ only in the value of nucleon effective mass, it generalizes well and it approximates the maximum star mass corresponding to the other model variants we consider with very high accuracy. This seems to indicate that the linear connection is a good approximation regardless of the bosonic interaction term used, it holds across different model variants. 
%
%

This observations is further supported by the connection between the radius of the maximum mass star ($R_{max\,M}$) and nuclear Landau mass. This connection is also present strong linear dependence and holds across all model variants we considered in this paper. The results are show on Fig.~\ref{fig:R_ml}, where the plotted linear relation is described by
%
%
\begin{equation}
R_{max\,M}= 21.94 - 0.015 \,  m_{L} \, .
\label{eq:M_ml_fit}
\end{equation}
%
%
applying $R$ is given in km units and $m_{L}$  is given in MeV. 

\begin{figure}[!h]
\begin{center}
\includegraphics[width=0.81\textwidth]{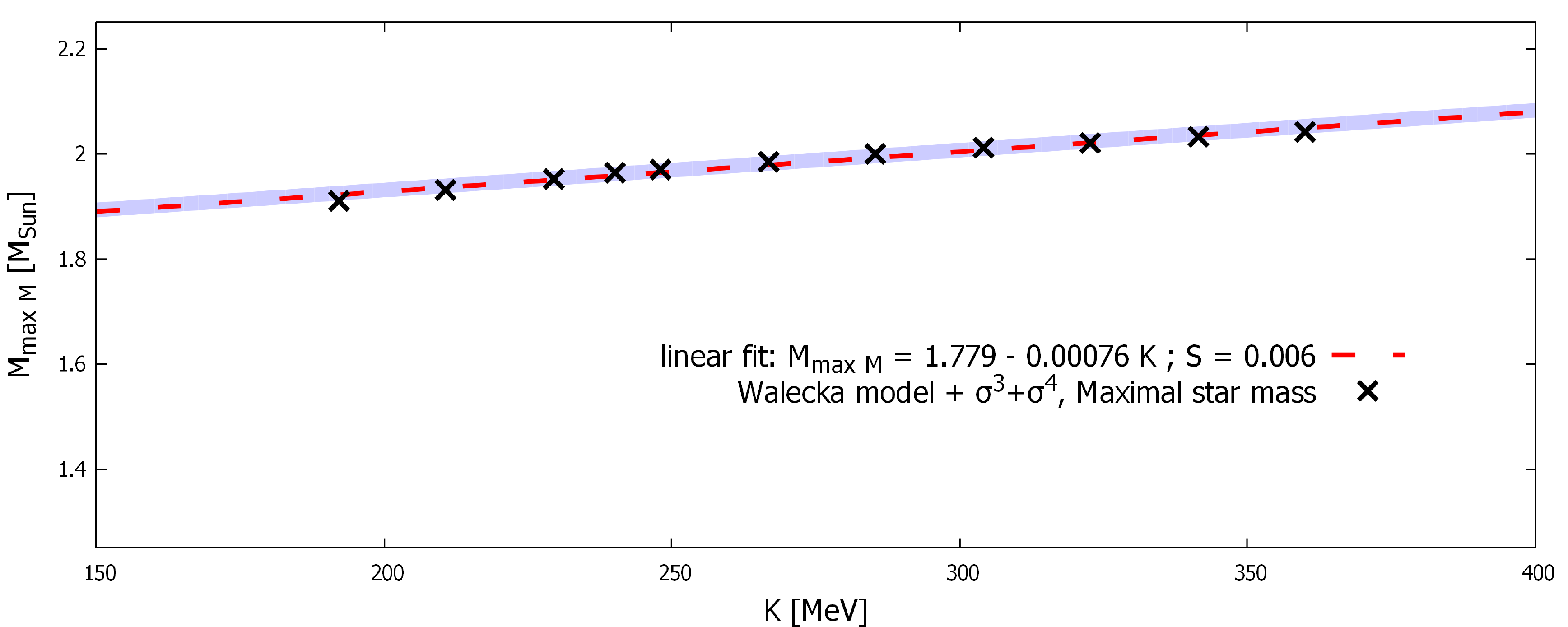}
\includegraphics[width=0.81\textwidth]{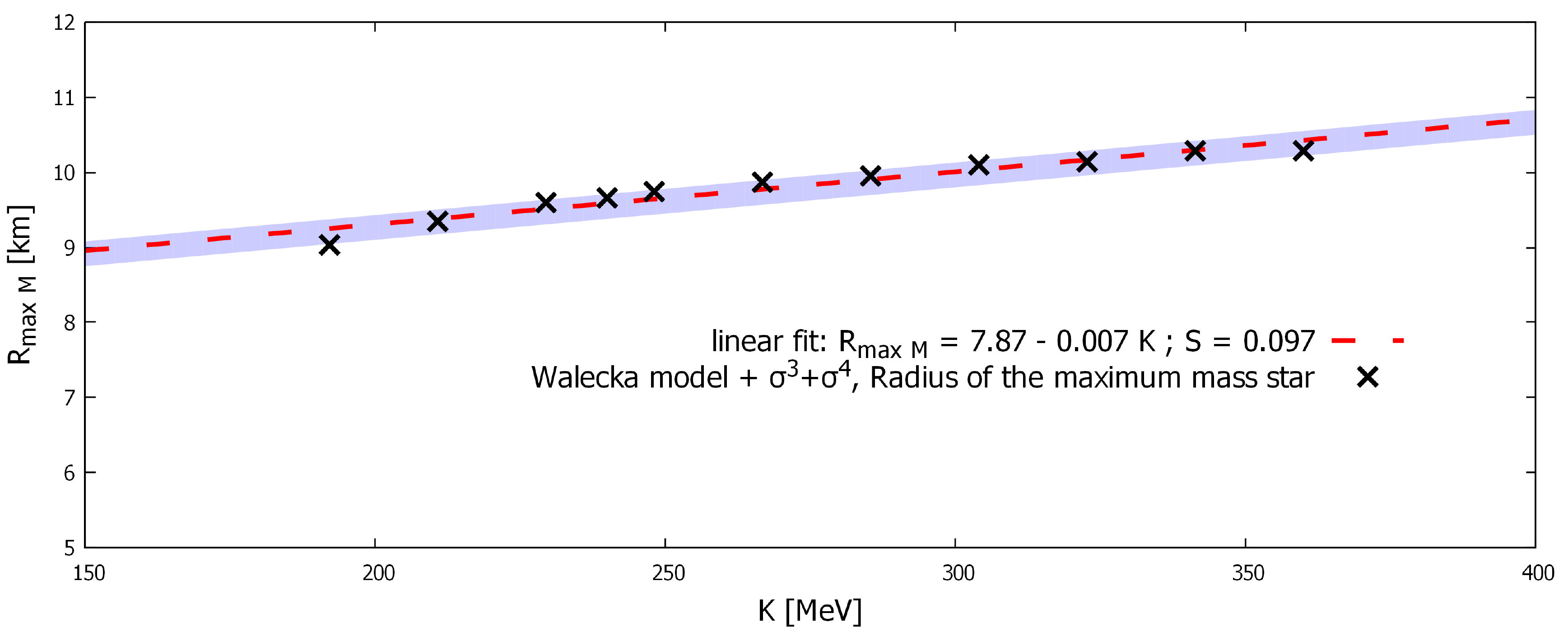}
\end{center}
\vspace*{-0.5 truecm}
\caption{\label{fig:M_K} The maximum mass star mass, $M_{Max\,M}$ ({\sl upper panel}) and radius, $R_{max\,M}$ ({\sl lower panel}) as function of the compression modulus, $K$. The dashed line is the best linear fit which is plotted as a guidance.}
\end{figure}

Since besides the nucleon Landau (effective) mass the only parameter which is different in our model variants is the incompressibility,  it is worth to study the relation between incompressibility and maximum star mass while keeping every other parameter constant. For this end in the model with $U_{34}$ interaction term, the Landau mass fixed for the value which is listed in Table~\ref{tab:fitting_data} and the model is solved to reproduce different values of incompressibility. All these different parametrizations gives a different dense matter equation of states which predict different $M$ - $R$ diagrams and different maximum mass star parameters, $M_{Max\,M}$. These results are summarized in the panels of Figure~\ref{fig:M_K}.
%
%

It can be seen in these plots that the the mass and radius of the maximum mass star are insesitive to the value of the incompressibility. The equations of the best (constant) linear fits are,
%
%
\begin{equation}
\begin{aligned}
M_{max\,M} &= 1.779 + 0.0008 \, K  \\
R_{max\,M} &= 7.870 + 0.0070 \, K \\
\end{aligned}
\label{eq:2 }
\end{equation}
where $M_{max\,M}$ is measured in units of $M_{Sun}$, $K$ in MeV units, and $R_{max\,M}$ is in km. 

The slope of both linear functions is tiny and even double the $K$ produces only a few percent of variation in the maximum mass star mass and radius. These results provide a heuristic understanding of the previous results on the sense matter properties: the maximum mass star radius and mass depends strongly on nucleon Landau mass, since it's dependence on incompressibility is negligible compared to how strongly it depends on the value of effective mass and all other parameters are kept constant.

\section{Summary}

We investigate the macroscopical observables of neutron stars with a symmetric dense nuclear and matter in the exterior. We compared three cases with different general bosonic interactions, we used  variations of the nuclear parameter fit method. We investigated how these bosonic interactions matter in the nuclear equation of state, and especially how the effective mass and Landau mass parameters play role in the nuclear potentials.

With variation of these microscopical parameters we explored the mass-radius groups in case of a Schwarzschild neutron star model.
We presented that the Landau mass of the nuclear matter have definite linear relation to the maximum mass and maximum radius of the neutron star, considered within the best $U_{34}$ model case. We also obtained that the effect of the incompressibility is negligible on these macroscopical parameters, with $M_{max\,M} \approx 1.78 $ and $R_{max\,M} = 7.87$ within the regime $K=$200-400, meaning $\lesssim 5\%$ variation for both cases.

These results support the idea, that evolution equation of the strongly interacting matter saves the magnitude of the uncertainty originated from the microscopical physical parameters, which appears as deterministic variation of the macroscopical observables as suggested in Refs.~\cite{Posfay:2017cor,Posfay:2016ygf}.

\funding{This work is supported by the Hungarian Research Fund (OTKA) under contracts No. K120660, K123815, NKM-81/2016 MTA-UA bilateral mobility program, and  PHAROS (CA16214), and THOR (CA15213) COST actions. 
Authors also acknowledges the computational resources for the Wigner GPU Laboratory.}

\reftitle{References}
\externalbibliography{yes}

\bibliography{walref07}

\begin{thebibliography}{-------}
\providecommand{\natexlab}[1]{#1}

\bibitem[NASA(2017)]{nasanicer}
NASA.
\newblock Nicer.
\newblock \url{https://www.nasa.gov/nicer}.

\bibitem[Merloni \em{et~al.}(2012)Merloni et~al.]{Merloni:2012uf}
Merloni, A.; others.
\newblock {eROSITA Science Book: Mapping the Structure of the Energetic
  Universe} {\bf 2012}.
\newblock  \href{http://xxx.lanl.gov/abs/1209.3114}{{\normalfont
  [arXiv:astro-ph.HE/1209.3114]}}.

\bibitem[Athena(2019)]{athena}
Athena.
\newblock The Athena X-ray Observatory.
\newblock \url{https://www.the-athena-x-ray-observatory.eu/}.

\bibitem[Ozel \em{et~al.}(2016)Ozel, Psaltis, Arzoumanian, Morsink, and
  Baubock]{Ozel:2015ykl}
Ozel, F.; Psaltis, D.; Arzoumanian, Z.; Morsink, S.; Baubock, M.
\newblock {Measuring Neutron Star Radii via Pulse Profile Modeling with NICER}.
\newblock {\em Astrophys. J.} {\bf 2016}, {\em 832},~92,
  \href{http://xxx.lanl.gov/abs/1512.03067}{{\normalfont
  [arXiv:astro-ph.HE/1512.03067]}}.
\newblock
  doi:{\changeurlcolor{black}\href{https://doi.org/10.3847/0004-637X/832/1/92}{\detokenize{10.3847/0004-637X/832/1/92}}}.

\bibitem[Watts(2019)]{Watts:2019lbs}
Watts, A.L.
\newblock {Constraining the neutron star equation of state using Pulse Profile
  Modeling}.
\newblock  {Xiamen-CUSTIPEN Workshop on the EOS of Dense Neutron-Rich Matter in
  the Era of Gravitational Wave Astronomy Xiamen, China, January 3-7, 2019},
  2019,  \href{http://xxx.lanl.gov/abs/1904.07012}{{\normalfont
  [arXiv:astro-ph.HE/1904.07012]}}.

\bibitem[Abbott \em{et~al.}(2018)Abbott, Abbott, Abbott, Acernese, others,
  Ackley, Adams, Adams, Addesso, Adhikari, Adya, Affeldt, Agarwal, Agathos,
  Agatsuma, Aggarwal, Aguiar, Aiello, Ain, Ajith, Allen, Allen, Allocca, Aloy,
  Altin, Amato, Ananyeva, Anderson, Anderson, Angelova, Antier, Appert, Arai,
  Araya, Areeda, Ar\`ene, Arnaud, Arun, Ascenzi, Ashton, Ast, Aston, Astone,
  Atallah, Aubin, Aufmuth, Aulbert, AultONeal, Austin, Avila-Alvarez, Babak,
  Bacon, Badaracco, Bader, Bae, Baker, Baldaccini, Ballardin, Ballmer,
  Banagiri, Barayoga, Barclay, Barish, Barker, Barkett, Barnum, Barone, Barr,
  Barsotti, Barsuglia, Barta, Bartlett, Bartos, Bassiri, Basti, Batch, Bawaj,
  Bayley, Bazzan, B\'ecsy, Beer, Bejger, Belahcene, Bell, Beniwal, Bensch,
  Berger, Bergmann, Bernuzzi, Bero, Berry, Bersanetti, Bertolini, Betzwieser,
  Bhandare, Bilenko, Bilgili, Billingsley, Billman, Birch, Birney, Birnholtz,
  Biscans, Biscoveanu, Bisht, Bitossi, Bizouard, Blackburn, Blackman, Blair,
  Blair, Blair, Bloemen, Bock, Bode, Boer, Boetzel, Bogaert, Bohe, Bondu,
  Bonilla, Bonnand, Booker, Boom, Booth, Bork, Boschi, Bose, Bossie, Bossilkov,
  Bosveld, Bouffanais, Bozzi, Bradaschia, Brady, Bramley, Branchesi, Brau,
  Briant, Brighenti, Brillet, Brinkmann, Brisson, Brockill, Brooks, Brown,
  Brunett, Buchanan, Buikema, Bulik, Bulten, Buonanno, Buskulic, Buy, Byer,
  Cabero, Cadonati, Cagnoli, Cahillane, Calder\'on~Bustillo, Callister,
  Calloni, Camp, Canepa, Canizares, Cannon, Cao, Cao, Capano, Capocasa,
  Carbognani, Caride, Carney, Carullo, Casanueva~Diaz, Casentini, Caudill,
  Cavagli\`a, Cavalier, Cavalieri, Cella, Cepeda, Cerd\'a-Dur\'an, Cerretani,
  Cesarini, Chaibi, Chamberlin, Chan, Chao, Charlton, Chase, Chassande-Mottin,
  Chatterjee, Chatziioannou, Cheeseboro, Chen, Chen, Chen, Cheng, Chia,
  Chincarini, Chiummo, Chmiel, Cho, Cho, Chow, Christensen, Chu, Chua, Chua,
  Chung, Chung, Ciani, Ciobanu, Ciolfi, Cipriano, Cirelli, Cirone, Clara,
  Clark, Clearwater, Cleva, Cocchieri, Coccia, Cohadon, Cohen, Colla, Collette,
  Collins, Cominsky, Constancio, Conti, Cooper, Corban, Corbitt,
  Cordero-Carri\'on, Corley, Cornish, Corsi, Cortese, Costa, Cotesta, Coughlin,
  Coughlin, Coulon, Countryman, Couvares, Covas, Cowan, Coward, Cowart, Coyne,
  Coyne, Creighton, Creighton, Cripe, Crowder, Cullen, Cumming, Cunningham,
  Cuoco, Canton, D\'alya, Danilishin, D'Antonio, Danzmann, Dasgupta,
  Da~Silva~Costa, Dattilo, Dave, Davier, Davis, Daw, Day, DeBra, Deenadayalan,
  Degallaix, De~Laurentis, Del\'eglise, Del~Pozzo, Demos, Denker, Dent,
  De~Pietri, Derby, Dergachev, De~Rosa, De~Rossi, DeSalvo, de~Varona,
  Dhurandhar, D\'{\i}az, Dietrich, Di~Fiore, Di~Giovanni, Di~Girolamo,
  Di~Lieto, Ding, Di~Pace, Di~Palma, Di~Renzo, Dmitriev, Doctor, Dolique,
  Donovan, Dooley, Doravari, Dorrington, Dovale~\'Alvarez, Downes, Drago,
  Dreissigacker, Driggers, Du, Dupej, Dwyer, Easter, Edo, Edwards, Effler,
  Eggenstein, Ehrens, Eichholz, Eikenberry, Eisenmann, Eisenstein, Essick,
  Estelles, Estevez, Etienne, Etzel, Evans, Evans, Fafone, Fair, Fairhurst,
  Fan, Farinon, Farr, Farr, Fauchon-Jones, Favata, Fays, Fee, Fehrmann, Feicht,
  Fejer, Feng, Fernandez-Galiana, Ferrante, Ferreira, Ferrini, Fidecaro, Fiori,
  Fiorucci, Fishbach, Fisher, Fishner, Fitz-Axen, Flaminio, Fletcher, Fong,
  Font, Forsyth, Forsyth, Fournier, Frasca, Frasconi, Frei, Freise, Frey, Frey,
  Fritschel, Frolov, Fulda, Fyffe, Gabbard, Gadre, Gaebel, Gair, Gammaitoni,
  Ganija, Gaonkar, Garcia, Garc\'{\i}a-Quir\'os, Garufi, Gateley, Gaudio, Gaur,
  Gayathri, Gemme, Genin, Gennai, George, George, Gergely, Germain, Ghonge,
  Ghosh, Ghosh, Ghosh, Giacomazzo, Giaime, Giardina, Giazotto, Gill, Giordano,
  Glover, Goetz, Goetz, Goncharov, Gonz\'alez, Gonzalez~Castro, Gopakumar,
  Gorodetsky, Gossan, Gosselin, Gouaty, Grado, Graef, Granata, Grant, Gras,
  Gray, Greco, Green, Green, Gretarsson, Groot, Grote, Grunewald, Gruning,
  Guidi, Gulati, Guo, Gupta, Gupta, Gushwa, Gustafson, Gustafson, Halim, Hall,
  Hall, Hamilton, Hamilton, Hammond, Haney, Hanke, Hanks, Hanna, Hannam,
  Hannuksela, Hanson, Hardwick, Harms, Harry, Harry, Hart, Haster, Haughian,
  Healy, Heidmann, Heintze, Heitmann, Hello, Hemming, Hendry, Heng, Hennig,
  Heptonstall, Hernandez, Heurs, Hild, Hinderer, Ho, Hoak, Hochheim, Hofman,
  Holland, Holt, Holz, Hopkins, Horst, Hough, Houston, Howell, Hreibi, Huerta,
  Huet, Hughey, Hulko, Husa, Huttner, Huynh-Dinh, Iess, Indik, Ingram, Inta,
  Intini, Irwin, Isa, Isac, Isi, Iyer, Izumi, Jacqmin, Jani, Jaranowski,
  Johnson, Johnson, Jones, Jones, Jonker, Ju, Junker, Kalaghatgi, Kalogera,
  Kamai, Kandhasamy, Kang, Kanner, Kapadia, Karki, Karvinen, Kasprzack,
  Katolik, Katsanevas, Katsavounidis, Katzman, Kaufer, Kawabe, Keerthana,
  K\'ef\'elian, Keitel, Kemball, Kennedy, Key, Khalili, Khamesra, Khan, Khan,
  Khan, Khan, Khazanov, Kijbunchoo, Kim, Kim, Kim, Kim, Kim, Kim, King, King,
  Kinley-Hanlon, Kirchhoff, Kissel, Kleybolte, Klimenko, Knowles, Koch,
  Koehlenbeck, Koley, Kondrashov, Kontos, Korobko, Korth, Kowalska, Kozak,
  Kr\"amer, Kringel, Krishnan, Kr\'olak, Kuehn, Kumar, Kumar, Kumar, Kuo,
  Kutynia, Kwang, Lackey, Lai, Landry, Landry, Lang, Lange, Lantz, Lanza,
  Lartaux-Vollard, Lasky, Laxen, Lazzarini, Lazzaro, Leaci, Leavey, Lee, Lee,
  Lee, Lee, Lee, Lehmann, Lenon, Leonardi, Leroy, Letendre, Levin, Li, Li, Li,
  Linker, Littenberg, Liu, Liu, Lo, Lockerbie, London, Longo, Lorenzini,
  Loriette, Lormand, Losurdo, Lough, Lousto, Lovelace, L\"uck, Lumaca,
  Lundgren, Lynch, Ma, Macas, Macfoy, Machenschalk, MacInnis, Macleod, Maga\~na
  Hernandez, Maga\~na Sandoval, Maga\~na Zertuche, Magee, Majorana, Maksimovic,
  Man, Mandic, Mangano, Mansell, Manske, Mantovani, Marchesoni, Marion,
  M\'arka, M\'arka, Markakis, Markosyan, Markowitz, Maros, Marquina, Martelli,
  Martellini, Martin, Martin, Martynov, Mason, Massera, Masserot, Massinger,
  Masso-Reid, Mastrogiovanni, Matas, Matichard, Matone, Mavalvala, Mazumder,
  McCann, McCarthy, McClelland, McCormick, McCuller, McGuire, McIver, McManus,
  McRae, McWilliams, Meacher, Meadors, Mehmet, Meidam, Mejuto-Villa, Melatos,
  Mendell, Mendoza-Gandara, Mercer, Mereni, Merilh, Merzougui, Meshkov,
  Messenger, Messick, Metzdorff, Meyers, Miao, Michel, Middleton, Mikhailov,
  Milano, Miller, Miller, Miller, Miller, Millhouse, Mills, Milovich-Goff,
  Minazzoli, Minenkov, Ming, Mishra, Mitra, Mitrofanov, Mitselmakher,
  Mittleman, Moffa, Mogushi, Mohan, Mohapatra, Montani, Moore, Moraru, Moreno,
  Morisaki, Mours, Mow-Lowry, Mueller, Muir, Mukherjee, Mukherjee, Mukherjee,
  Mukund, Mullavey, Munch, Mu\~niz, Muratore, Murray, Nagar, Napier,
  Nardecchia, Naticchioni, Nayak, Neilson, Nelemans, Nelson, Nery, Neunzert,
  Nevin, Newport, Ng, Ng, Nguyen, Nguyen, Nichols, Nielsen, Nissanke, Nitz,
  Nocera, Nolting, North, Nuttall, Obergaulinger, Oberling, O'Brien, O'Dea,
  Ogin, Oh, Oh, Ohme, Ohta, Okada, Oliver, Oppermann, Oram, O'Reilly, Ormiston,
  Ortega, O'Shaughnessy, Ossokine, Ottaway, Overmier, Owen, Pace, Pagano, Page,
  Page, Pai, Pai, Palamos, Palashov, Palomba, Pal-Singh, Pan, Pan, Pang, Pang,
  Pankow, Pannarale, Pant, Paoletti, Paoli, Papa, Parida, Parker, Pascucci,
  Pasqualetti, Passaquieti, Passuello, Patil, Patricelli, Pearlstone, Pedersen,
  Pedraza, Pedurand, Pekowsky, Pele, Penn, Perego, Perez, Perreca, Perri,
  Pfeiffer, Phelps, Phukon, Piccinni, Pichot, Piergiovanni, Pierro, Pillant,
  Pinard, Pinto, Pirello, Pitkin, Poggiani, Popolizio, Porter, Possenti, Post,
  Powell, Prasad, Pratt, Pratten, Predoi, Prestegard, Principe, Privitera,
  Prodi, Prokhorov, Puncken, Punturo, Puppo, P\"urrer, Qi, Quetschke, Quintero,
  Quitzow-James, Raab, Rabeling, Radkins, Raffai, Raja, Rajan, Rajbhandari,
  Rakhmanov, Ramirez, Ramos-Buades, Rana, Rapagnani, Raymond, Razzano, Read,
  Regimbau, Rei, Reid, Reitze, Ren, Ricci, Ricker, Riemenschneider, Riles,
  Rizzo, Robertson, Robie, Robinet, Robson, Rocchi, Rolland, Rollins, Roma,
  Romano, Romel, Romie, Rosi\ifmmode~\acute{n}\else \'{n}\fi{}ska, Ross, Rowan,
  R\"udiger, Ruggi, Rutins, Ryan, Sachdev, Sadecki, Sakellariadou, Salconi,
  Saleem, Salemi, Samajdar, Sammut, Sampson, Sanchez, Sanchez, Sanchis-Gual,
  Sandberg, Sanders, Sarin, Sassolas, Sathyaprakash, Saulson, Sauter, Savage,
  Sawadsky, Schale, Scheel, Scheuer, Schmidt, Schnabel, Schofield, Sch\"onbeck,
  Schreiber, Schuette, Schulte, Schutz, Schwalbe, Scott, Scott, Seidel,
  Sellers, Sengupta, Sentenac, Sequino, Sergeev, Setyawati, Shaddock, Shaffer,
  Shah, Shahriar, Shaner, Shao, Shapiro, Shawhan, Shen, Shoemaker, Shoemaker,
  Siellez, Siemens, Sieniawska, Sigg, Silva, Singer, Singh, Singhal, Sintes,
  Slagmolen, Slaven-Blair, Smith, Smith, Smith, Somala, Son, Sorazu,
  Sorrentino, Souradeep, Spencer, Srivastava, Staats, Steinke, Steinlechner,
  Steinlechner, Steinmeyer, Steltner, Stevenson, Stocks, Stone, Stops, Strain,
  Stratta, Strigin, Strunk, Sturani, Stuver, Summerscales, Sun, Sunil, Suresh,
  Sutton, Swinkels, Szczepa\ifmmode~\acute{n}\else \'{n}\fi{}czyk, Tacca, Tait,
  Talbot, Talukder, Tanner, T\'apai, Taracchini, Tasson, Taylor, Taylor,
  Tewari, Theeg, Thies, Thomas, Thomas, Thomas, Thorne, Thrane, Tiwari, Tiwari,
  Tokmakov, Toland, Tonelli, Tornasi, Torres-Forn\'e, Torrie, T\"oyr\"a,
  Travasso, Traylor, Trinastic, Tringali, Trovato, Trozzo, Tsang, Tse, Tso,
  Tsuna, Tsukada, Tuyenbayev, Ueno, Ugolini, Urban, Usman, Vahlbruch, Vajente,
  Valdes, van Bakel, van Beuzekom, van~den Brand, Van Den~Broeck, Vander-Hyde,
  van~der Schaaf, van Heijningen, van Veggel, Vardaro, Varma, Vass, Vas\'uth,
  Vecchio, Vedovato, Veitch, Veitch, Venkateswara, Venugopalan, Verkindt,
  Vetrano, Vicer\'e, Viets, Vinciguerra, Vine, Vinet, Vitale, Vo, Vocca,
  Vorvick, Vyatchanin, Wade, Wade, Wade, Walet, Walker, Wallace, Walsh, Wang,
  Wang, Wang, Wang, Wang, Ward, Warner, Was, Watchi, Weaver, Wei, Weinert,
  Weinstein, Weiss, Wellmann, Wen, Wessel, We\ss{}els, Westerweck, Wette,
  Whelan, Whiting, Whittle, Wilken, Williams, Williams, Williamson, Willis,
  Willke, Wimmer, Winkler, Wipf, Wittel, Woan, Woehler, Wofford, Wong, Worden,
  Wright, Wu, Wysocki, Xiao, Yam, Yamamoto, Yancey, Yang, Yap, Yazback, Yu, Yu,
  Yvert, Zadro\ifmmode~\dot{z}\else \.{z}\fi{}ny, Zanolin, Zelenova, Zendri,
  Zevin, Zhang, Zhang, Zhang, Zhang, Zhang, Zhao, Zhou, Zhou, Zhu, Zhu,
  Zimmerman, Zlochower, Zucker, and Zweizig]{Abott_gw1708_radius}
Abbott, B.P.; Abbott, R.; Abbott, T.D.; Acernese, F.; others.; Ackley, K.;
  Adams, C.; Adams, T.; Addesso, P.; Adhikari, R.X.; Adya, V.B.; Affeldt, C.;
  Agarwal, B.; Agathos, M.; Agatsuma, K.; Aggarwal, N.; Aguiar, O.D.; Aiello,
  L.; Ain, A.; Ajith, P.; Allen, B.; Allen, G.; Allocca, A.; Aloy, M.A.; Altin,
  P.A.; Amato, A.; Ananyeva, A.; Anderson, S.B.; Anderson, W.G.; Angelova,
  S.V.; Antier, S.; Appert, S.; Arai, K.; Araya, M.C.; Areeda, J.S.; Ar\`ene,
  M.; Arnaud, N.; Arun, K.G.; Ascenzi, S.; Ashton, G.; Ast, M.; Aston, S.M.;
  Astone, P.; Atallah, D.V.; Aubin, F.; Aufmuth, P.; Aulbert, C.; AultONeal,
  K.; Austin, C.; Avila-Alvarez, A.; Babak, S.; Bacon, P.; Badaracco, F.;
  Bader, M.K.M.; Bae, S.; Baker, P.T.; Baldaccini, F.; Ballardin, G.; Ballmer,
  S.W.; Banagiri, S.; Barayoga, J.C.; Barclay, S.E.; Barish, B.C.; Barker, D.;
  Barkett, K.; Barnum, S.; Barone, F.; Barr, B.; Barsotti, L.; Barsuglia, M.;
  Barta, D.; Bartlett, J.; Bartos, I.; Bassiri, R.; Basti, A.; Batch, J.C.;
  Bawaj, M.; Bayley, J.C.; Bazzan, M.; B\'ecsy, B.; Beer, C.; Bejger, M.;
  Belahcene, I.; Bell, A.S.; Beniwal, D.; Bensch, M.; Berger, B.K.; Bergmann,
  G.; Bernuzzi, S.; Bero, J.J.; Berry, C.P.L.; Bersanetti, D.; Bertolini, A.;
  Betzwieser, J.; Bhandare, R.; Bilenko, I.A.; Bilgili, S.A.; Billingsley, G.;
  Billman, C.R.; Birch, J.; Birney, R.; Birnholtz, O.; Biscans, S.; Biscoveanu,
  S.; Bisht, A.; Bitossi, M.; Bizouard, M.A.; Blackburn, J.K.; Blackman, J.;
  Blair, C.D.; Blair, D.G.; Blair, R.M.; Bloemen, S.; Bock, O.; Bode, N.; Boer,
  M.; Boetzel, Y.; Bogaert, G.; Bohe, A.; Bondu, F.; Bonilla, E.; Bonnand, R.;
  Booker, P.; Boom, B.A.; Booth, C.D.; Bork, R.; Boschi, V.; Bose, S.; Bossie,
  K.; Bossilkov, V.; Bosveld, J.; Bouffanais, Y.; Bozzi, A.; Bradaschia, C.;
  Brady, P.R.; Bramley, A.; Branchesi, M.; Brau, J.E.; Briant, T.; Brighenti,
  F.; Brillet, A.; Brinkmann, M.; Brisson, V.; Brockill, P.; Brooks, A.F.;
  Brown, D.D.; Brunett, S.; Buchanan, C.C.; Buikema, A.; Bulik, T.; Bulten,
  H.J.; Buonanno, A.; Buskulic, D.; Buy, C.; Byer, R.L.; Cabero, M.; Cadonati,
  L.; Cagnoli, G.; Cahillane, C.; Calder\'on~Bustillo, J.; Callister, T.A.;
  Calloni, E.; Camp, J.B.; Canepa, M.; Canizares, P.; Cannon, K.C.; Cao, H.;
  Cao, J.; Capano, C.D.; Capocasa, E.; Carbognani, F.; Caride, S.; Carney,
  M.F.; Carullo, G.; Casanueva~Diaz, J.; Casentini, C.; Caudill, S.;
  Cavagli\`a, M.; Cavalier, F.; Cavalieri, R.; Cella, G.; Cepeda, C.B.;
  Cerd\'a-Dur\'an, P.; Cerretani, G.; Cesarini, E.; Chaibi, O.; Chamberlin,
  S.J.; Chan, M.; Chao, S.; Charlton, P.; Chase, E.; Chassande-Mottin, E.;
  Chatterjee, D.; Chatziioannou, K.; Cheeseboro, B.D.; Chen, H.Y.; Chen, X.;
  Chen, Y.; Cheng, H.P.; Chia, H.Y.; Chincarini, A.; Chiummo, A.; Chmiel, T.;
  Cho, H.S.; Cho, M.; Chow, J.H.; Christensen, N.; Chu, Q.; Chua, A.J.K.; Chua,
  S.; Chung, K.W.; Chung, S.; Ciani, G.; Ciobanu, A.A.; Ciolfi, R.; Cipriano,
  F.; Cirelli, C.E.; Cirone, A.; Clara, F.; Clark, J.A.; Clearwater, P.; Cleva,
  F.; Cocchieri, C.; Coccia, E.; Cohadon, P.F.; Cohen, D.; Colla, A.; Collette,
  C.G.; Collins, C.; Cominsky, L.R.; Constancio, M.; Conti, L.; Cooper, S.J.;
  Corban, P.; Corbitt, T.R.; Cordero-Carri\'on, I.; Corley, K.R.; Cornish, N.;
  Corsi, A.; Cortese, S.; Costa, C.A.; Cotesta, R.; Coughlin, M.W.; Coughlin,
  S.B.; Coulon, J.P.; Countryman, S.T.; Couvares, P.; Covas, P.B.; Cowan, E.E.;
  Coward, D.M.; Cowart, M.J.; Coyne, D.C.; Coyne, R.; Creighton, J.D.E.;
  Creighton, T.D.; Cripe, J.; Crowder, S.G.; Cullen, T.J.; Cumming, A.;
  Cunningham, L.; Cuoco, E.; Canton, T.D.; D\'alya, G.; Danilishin, S.L.;
  D'Antonio, S.; Danzmann, K.; Dasgupta, A.; Da~Silva~Costa, C.F.; Dattilo, V.;
  Dave, I.; Davier, M.; Davis, D.; Daw, E.J.; Day, B.; DeBra, D.; Deenadayalan,
  M.; Degallaix, J.; De~Laurentis, M.; Del\'eglise, S.; Del~Pozzo, W.; Demos,
  N.; Denker, T.; Dent, T.; De~Pietri, R.; Derby, J.; Dergachev, V.; De~Rosa,
  R.; De~Rossi, C.; DeSalvo, R.; de~Varona, O.; Dhurandhar, S.; D\'{\i}az,
  M.C.; Dietrich, T.; Di~Fiore, L.; Di~Giovanni, M.; Di~Girolamo, T.; Di~Lieto,
  A.; Ding, B.; Di~Pace, S.; Di~Palma, I.; Di~Renzo, F.; Dmitriev, A.; Doctor,
  Z.; Dolique, V.; Donovan, F.; Dooley, K.L.; Doravari, S.; Dorrington, I.;
  Dovale~\'Alvarez, M.; Downes, T.P.; Drago, M.; Dreissigacker, C.; Driggers,
  J.C.; Du, Z.; Dupej, P.; Dwyer, S.E.; Easter, P.J.; Edo, T.B.; Edwards, M.C.;
  Effler, A.; Eggenstein, H.B.; Ehrens, P.; Eichholz, J.; Eikenberry, S.S.;
  Eisenmann, M.; Eisenstein, R.A.; Essick, R.C.; Estelles, H.; Estevez, D.;
  Etienne, Z.B.; Etzel, T.; Evans, M.; Evans, T.M.; Fafone, V.; Fair, H.;
  Fairhurst, S.; Fan, X.; Farinon, S.; Farr, B.; Farr, W.M.; Fauchon-Jones,
  E.J.; Favata, M.; Fays, M.; Fee, C.; Fehrmann, H.; Feicht, J.; Fejer, M.M.;
  Feng, F.; Fernandez-Galiana, A.; Ferrante, I.; Ferreira, E.C.; Ferrini, F.;
  Fidecaro, F.; Fiori, I.; Fiorucci, D.; Fishbach, M.; Fisher, R.P.; Fishner,
  J.M.; Fitz-Axen, M.; Flaminio, R.; Fletcher, M.; Fong, H.; Font, J.A.;
  Forsyth, P.W.F.; Forsyth, S.S.; Fournier, J.D.; Frasca, S.; Frasconi, F.;
  Frei, Z.; Freise, A.; Frey, R.; Frey, V.; Fritschel, P.; Frolov, V.V.; Fulda,
  P.; Fyffe, M.; Gabbard, H.A.; Gadre, B.U.; Gaebel, S.M.; Gair, J.R.;
  Gammaitoni, L.; Ganija, M.R.; Gaonkar, S.G.; Garcia, A.;
  Garc\'{\i}a-Quir\'os, C.; Garufi, F.; Gateley, B.; Gaudio, S.; Gaur, G.;
  Gayathri, V.; Gemme, G.; Genin, E.; Gennai, A.; George, D.; George, J.;
  Gergely, L.; Germain, V.; Ghonge, S.; Ghosh, A.; Ghosh, A.; Ghosh, S.;
  Giacomazzo, B.; Giaime, J.A.; Giardina, K.D.; Giazotto, A.; Gill, K.;
  Giordano, G.; Glover, L.; Goetz, E.; Goetz, R.; Goncharov, B.; Gonz\'alez,
  G.; Gonzalez~Castro, J.M.; Gopakumar, A.; Gorodetsky, M.L.; Gossan, S.E.;
  Gosselin, M.; Gouaty, R.; Grado, A.; Graef, C.; Granata, M.; Grant, A.; Gras,
  S.; Gray, C.; Greco, G.; Green, A.C.; Green, R.; Gretarsson, E.M.; Groot, P.;
  Grote, H.; Grunewald, S.; Gruning, P.; Guidi, G.M.; Gulati, H.K.; Guo, X.;
  Gupta, A.; Gupta, M.K.; Gushwa, K.E.; Gustafson, E.K.; Gustafson, R.; Halim,
  O.; Hall, B.R.; Hall, E.D.; Hamilton, E.Z.; Hamilton, H.F.; Hammond, G.;
  Haney, M.; Hanke, M.M.; Hanks, J.; Hanna, C.; Hannam, M.D.; Hannuksela, O.A.;
  Hanson, J.; Hardwick, T.; Harms, J.; Harry, G.M.; Harry, I.W.; Hart, M.J.;
  Haster, C.J.; Haughian, K.; Healy, J.; Heidmann, A.; Heintze, M.C.; Heitmann,
  H.; Hello, P.; Hemming, G.; Hendry, M.; Heng, I.S.; Hennig, J.; Heptonstall,
  A.W.; Hernandez, F.J.; Heurs, M.; Hild, S.; Hinderer, T.; Ho, W.C.G.; Hoak,
  D.; Hochheim, S.; Hofman, D.; Holland, N.A.; Holt, K.; Holz, D.E.; Hopkins,
  P.; Horst, C.; Hough, J.; Houston, E.A.; Howell, E.J.; Hreibi, A.; Huerta,
  E.A.; Huet, D.; Hughey, B.; Hulko, M.; Husa, S.; Huttner, S.H.; Huynh-Dinh,
  T.; Iess, A.; Indik, N.; Ingram, C.; Inta, R.; Intini, G.; Irwin, B.S.; Isa,
  H.N.; Isac, J.M.; Isi, M.; Iyer, B.R.; Izumi, K.; Jacqmin, T.; Jani, K.;
  Jaranowski, P.; Johnson, D.S.; Johnson, W.W.; Jones, D.I.; Jones, R.; Jonker,
  R.J.G.; Ju, L.; Junker, J.; Kalaghatgi, C.V.; Kalogera, V.; Kamai, B.;
  Kandhasamy, S.; Kang, G.; Kanner, J.B.; Kapadia, S.J.; Karki, S.; Karvinen,
  K.S.; Kasprzack, M.; Katolik, M.; Katsanevas, S.; Katsavounidis, E.; Katzman,
  W.; Kaufer, S.; Kawabe, K.; Keerthana, N.V.; K\'ef\'elian, F.; Keitel, D.;
  Kemball, A.J.; Kennedy, R.; Key, J.S.; Khalili, F.Y.; Khamesra, B.; Khan, H.;
  Khan, I.; Khan, S.; Khan, Z.; Khazanov, E.A.; Kijbunchoo, N.; Kim, C.; Kim,
  J.C.; Kim, K.; Kim, W.; Kim, W.S.; Kim, Y.M.; King, E.J.; King, P.J.;
  Kinley-Hanlon, M.; Kirchhoff, R.; Kissel, J.S.; Kleybolte, L.; Klimenko, S.;
  Knowles, T.D.; Koch, P.; Koehlenbeck, S.M.; Koley, S.; Kondrashov, V.;
  Kontos, A.; Korobko, M.; Korth, W.Z.; Kowalska, I.; Kozak, D.B.; Kr\"amer,
  C.; Kringel, V.; Krishnan, B.; Kr\'olak, A.; Kuehn, G.; Kumar, P.; Kumar, R.;
  Kumar, S.; Kuo, L.; Kutynia, A.; Kwang, S.; Lackey, B.D.; Lai, K.H.; Landry,
  M.; Landry, P.; Lang, R.N.; Lange, J.; Lantz, B.; Lanza, R.K.;
  Lartaux-Vollard, A.; Lasky, P.D.; Laxen, M.; Lazzarini, A.; Lazzaro, C.;
  Leaci, P.; Leavey, S.; Lee, C.H.; Lee, H.K.; Lee, H.M.; Lee, H.W.; Lee, K.;
  Lehmann, J.; Lenon, A.; Leonardi, M.; Leroy, N.; Letendre, N.; Levin, Y.; Li,
  J.; Li, T.G.F.; Li, X.; Linker, S.D.; Littenberg, T.B.; Liu, J.; Liu, X.; Lo,
  R.K.L.; Lockerbie, N.A.; London, L.T.; Longo, A.; Lorenzini, M.; Loriette,
  V.; Lormand, M.; Losurdo, G.; Lough, J.D.; Lousto, C.O.; Lovelace, G.;
  L\"uck, H.; Lumaca, D.; Lundgren, A.P.; Lynch, R.; Ma, Y.; Macas, R.; Macfoy,
  S.; Machenschalk, B.; MacInnis, M.; Macleod, D.M.; Maga\~na Hernandez, I.;
  Maga\~na Sandoval, F.; Maga\~na Zertuche, L.; Magee, R.M.; Majorana, E.;
  Maksimovic, I.; Man, N.; Mandic, V.; Mangano, V.; Mansell, G.L.; Manske, M.;
  Mantovani, M.; Marchesoni, F.; Marion, F.; M\'arka, S.; M\'arka, Z.;
  Markakis, C.; Markosyan, A.S.; Markowitz, A.; Maros, E.; Marquina, A.;
  Martelli, F.; Martellini, L.; Martin, I.W.; Martin, R.M.; Martynov, D.V.;
  Mason, K.; Massera, E.; Masserot, A.; Massinger, T.J.; Masso-Reid, M.;
  Mastrogiovanni, S.; Matas, A.; Matichard, F.; Matone, L.; Mavalvala, N.;
  Mazumder, N.; McCann, J.J.; McCarthy, R.; McClelland, D.E.; McCormick, S.;
  McCuller, L.; McGuire, S.C.; McIver, J.; McManus, D.J.; McRae, T.;
  McWilliams, S.T.; Meacher, D.; Meadors, G.D.; Mehmet, M.; Meidam, J.;
  Mejuto-Villa, E.; Melatos, A.; Mendell, G.; Mendoza-Gandara, D.; Mercer,
  R.A.; Mereni, L.; Merilh, E.L.; Merzougui, M.; Meshkov, S.; Messenger, C.;
  Messick, C.; Metzdorff, R.; Meyers, P.M.; Miao, H.; Michel, C.; Middleton,
  H.; Mikhailov, E.E.; Milano, L.; Miller, A.L.; Miller, A.; Miller, B.B.;
  Miller, J.; Millhouse, M.; Mills, J.; Milovich-Goff, M.C.; Minazzoli, O.;
  Minenkov, Y.; Ming, J.; Mishra, C.; Mitra, S.; Mitrofanov, V.P.;
  Mitselmakher, G.; Mittleman, R.; Moffa, D.; Mogushi, K.; Mohan, M.;
  Mohapatra, S.R.P.; Montani, M.; Moore, C.J.; Moraru, D.; Moreno, G.;
  Morisaki, S.; Mours, B.; Mow-Lowry, C.M.; Mueller, G.; Muir, A.W.; Mukherjee,
  A.; Mukherjee, D.; Mukherjee, S.; Mukund, N.; Mullavey, A.; Munch, J.;
  Mu\~niz, E.A.; Muratore, M.; Murray, P.G.; Nagar, A.; Napier, K.; Nardecchia,
  I.; Naticchioni, L.; Nayak, R.K.; Neilson, J.; Nelemans, G.; Nelson, T.J.N.;
  Nery, M.; Neunzert, A.; Nevin, L.; Newport, J.M.; Ng, K.Y.; Ng, S.; Nguyen,
  P.; Nguyen, T.T.; Nichols, D.; Nielsen, A.B.; Nissanke, S.; Nitz, A.; Nocera,
  F.; Nolting, D.; North, C.; Nuttall, L.K.; Obergaulinger, M.; Oberling, J.;
  O'Brien, B.D.; O'Dea, G.D.; Ogin, G.H.; Oh, J.J.; Oh, S.H.; Ohme, F.; Ohta,
  H.; Okada, M.A.; Oliver, M.; Oppermann, P.; Oram, R.J.; O'Reilly, B.;
  Ormiston, R.; Ortega, L.F.; O'Shaughnessy, R.; Ossokine, S.; Ottaway, D.J.;
  Overmier, H.; Owen, B.J.; Pace, A.E.; Pagano, G.; Page, J.; Page, M.A.; Pai,
  A.; Pai, S.A.; Palamos, J.R.; Palashov, O.; Palomba, C.; Pal-Singh, A.; Pan,
  H.; Pan, H.W.; Pang, B.; Pang, P.T.H.; Pankow, C.; Pannarale, F.; Pant, B.C.;
  Paoletti, F.; Paoli, A.; Papa, M.A.; Parida, A.; Parker, W.; Pascucci, D.;
  Pasqualetti, A.; Passaquieti, R.; Passuello, D.; Patil, M.; Patricelli, B.;
  Pearlstone, B.L.; Pedersen, C.; Pedraza, M.; Pedurand, R.; Pekowsky, L.;
  Pele, A.; Penn, S.; Perego, A.; Perez, C.J.; Perreca, A.; Perri, L.M.;
  Pfeiffer, H.P.; Phelps, M.; Phukon, K.S.; Piccinni, O.J.; Pichot, M.;
  Piergiovanni, F.; Pierro, V.; Pillant, G.; Pinard, L.; Pinto, I.M.; Pirello,
  M.; Pitkin, M.; Poggiani, R.; Popolizio, P.; Porter, E.K.; Possenti, L.;
  Post, A.; Powell, J.; Prasad, J.; Pratt, J.W.W.; Pratten, G.; Predoi, V.;
  Prestegard, T.; Principe, M.; Privitera, S.; Prodi, G.A.; Prokhorov, L.G.;
  Puncken, O.; Punturo, M.; Puppo, P.; P\"urrer, M.; Qi, H.; Quetschke, V.;
  Quintero, E.A.; Quitzow-James, R.; Raab, F.J.; Rabeling, D.S.; Radkins, H.;
  Raffai, P.; Raja, S.; Rajan, C.; Rajbhandari, B.; Rakhmanov, M.; Ramirez,
  K.E.; Ramos-Buades, A.; Rana, J.; Rapagnani, P.; Raymond, V.; Razzano, M.;
  Read, J.; Regimbau, T.; Rei, L.; Reid, S.; Reitze, D.H.; Ren, W.; Ricci, F.;
  Ricker, P.M.; Riemenschneider, G.M.; Riles, K.; Rizzo, M.; Robertson, N.A.;
  Robie, R.; Robinet, F.; Robson, T.; Rocchi, A.; Rolland, L.; Rollins, J.G.;
  Roma, V.J.; Romano, R.; Romel, C.L.; Romie, J.H.; Rosi\ifmmode~\acute{n}\else
  \'{n}\fi{}ska, D.; Ross, M.P.; Rowan, S.; R\"udiger, A.; Ruggi, P.; Rutins,
  G.; Ryan, K.; Sachdev, S.; Sadecki, T.; Sakellariadou, M.; Salconi, L.;
  Saleem, M.; Salemi, F.; Samajdar, A.; Sammut, L.; Sampson, L.M.; Sanchez,
  E.J.; Sanchez, L.E.; Sanchis-Gual, N.; Sandberg, V.; Sanders, J.R.; Sarin,
  N.; Sassolas, B.; Sathyaprakash, B.S.; Saulson, P.R.; Sauter, O.; Savage,
  R.L.; Sawadsky, A.; Schale, P.; Scheel, M.; Scheuer, J.; Schmidt, P.;
  Schnabel, R.; Schofield, R.M.S.; Sch\"onbeck, A.; Schreiber, E.; Schuette,
  D.; Schulte, B.W.; Schutz, B.F.; Schwalbe, S.G.; Scott, J.; Scott, S.M.;
  Seidel, E.; Sellers, D.; Sengupta, A.S.; Sentenac, D.; Sequino, V.; Sergeev,
  A.; Setyawati, Y.; Shaddock, D.A.; Shaffer, T.J.; Shah, A.A.; Shahriar, M.S.;
  Shaner, M.B.; Shao, L.; Shapiro, B.; Shawhan, P.; Shen, H.; Shoemaker, D.H.;
  Shoemaker, D.M.; Siellez, K.; Siemens, X.; Sieniawska, M.; Sigg, D.; Silva,
  A.D.; Singer, L.P.; Singh, A.; Singhal, A.; Sintes, A.M.; Slagmolen, B.J.J.;
  Slaven-Blair, T.J.; Smith, B.; Smith, J.R.; Smith, R.J.E.; Somala, S.; Son,
  E.J.; Sorazu, B.; Sorrentino, F.; Souradeep, T.; Spencer, A.P.; Srivastava,
  A.K.; Staats, K.; Steinke, M.; Steinlechner, J.; Steinlechner, S.;
  Steinmeyer, D.; Steltner, B.; Stevenson, S.P.; Stocks, D.; Stone, R.; Stops,
  D.J.; Strain, K.A.; Stratta, G.; Strigin, S.E.; Strunk, A.; Sturani, R.;
  Stuver, A.L.; Summerscales, T.Z.; Sun, L.; Sunil, S.; Suresh, J.; Sutton,
  P.J.; Swinkels, B.L.; Szczepa\ifmmode~\acute{n}\else \'{n}\fi{}czyk, M.J.;
  Tacca, M.; Tait, S.C.; Talbot, C.; Talukder, D.; Tanner, D.B.; T\'apai, M.;
  Taracchini, A.; Tasson, J.D.; Taylor, J.A.; Taylor, R.; Tewari, S.V.; Theeg,
  T.; Thies, F.; Thomas, E.G.; Thomas, M.; Thomas, P.; Thorne, K.A.; Thrane,
  E.; Tiwari, S.; Tiwari, V.; Tokmakov, K.V.; Toland, K.; Tonelli, M.; Tornasi,
  Z.; Torres-Forn\'e, A.; Torrie, C.I.; T\"oyr\"a, D.; Travasso, F.; Traylor,
  G.; Trinastic, J.; Tringali, M.C.; Trovato, A.; Trozzo, L.; Tsang, K.W.; Tse,
  M.; Tso, R.; Tsuna, D.; Tsukada, L.; Tuyenbayev, D.; Ueno, K.; Ugolini, D.;
  Urban, A.L.; Usman, S.A.; Vahlbruch, H.; Vajente, G.; Valdes, G.; van Bakel,
  N.; van Beuzekom, M.; van~den Brand, J.F.J.; Van Den~Broeck, C.; Vander-Hyde,
  D.C.; van~der Schaaf, L.; van Heijningen, J.V.; van Veggel, A.A.; Vardaro,
  M.; Varma, V.; Vass, S.; Vas\'uth, M.; Vecchio, A.; Vedovato, G.; Veitch, J.;
  Veitch, P.J.; Venkateswara, K.; Venugopalan, G.; Verkindt, D.; Vetrano, F.;
  Vicer\'e, A.; Viets, A.D.; Vinciguerra, S.; Vine, D.J.; Vinet, J.Y.; Vitale,
  S.; Vo, T.; Vocca, H.; Vorvick, C.; Vyatchanin, S.P.; Wade, A.R.; Wade, L.E.;
  Wade, M.; Walet, R.; Walker, M.; Wallace, L.; Walsh, S.; Wang, G.; Wang, H.;
  Wang, J.Z.; Wang, W.H.; Wang, Y.F.; Ward, R.L.; Warner, J.; Was, M.; Watchi,
  J.; Weaver, B.; Wei, L.W.; Weinert, M.; Weinstein, A.J.; Weiss, R.; Wellmann,
  F.; Wen, L.; Wessel, E.K.; We\ss{}els, P.; Westerweck, J.; Wette, K.; Whelan,
  J.T.; Whiting, B.F.; Whittle, C.; Wilken, D.; Williams, D.; Williams, R.D.;
  Williamson, A.R.; Willis, J.L.; Willke, B.; Wimmer, M.H.; Winkler, W.; Wipf,
  C.C.; Wittel, H.; Woan, G.; Woehler, J.; Wofford, J.K.; Wong, W.K.; Worden,
  J.; Wright, J.L.; Wu, D.S.; Wysocki, D.M.; Xiao, S.; Yam, W.; Yamamoto, H.;
  Yancey, C.C.; Yang, L.; Yap, M.J.; Yazback, M.; Yu, H.; Yu, H.; Yvert, M.;
  Zadro\ifmmode~\dot{z}\else \.{z}\fi{}ny, A.; Zanolin, M.; Zelenova, T.;
  Zendri, J.P.; Zevin, M.; Zhang, J.; Zhang, L.; Zhang, M.; Zhang, T.; Zhang,
  Y.H.; Zhao, C.; Zhou, M.; Zhou, Z.; Zhu, S.J.; Zhu, X.J.; Zimmerman, A.B.;
  Zlochower, Y.; Zucker, M.E.; Zweizig, J.
\newblock GW170817: Measurements of Neutron Star Radii and Equation of State.
\newblock {\em Phys. Rev. Lett.} {\bf 2018}, {\em 121},~161101.
\newblock
  doi:{\changeurlcolor{black}\href{https://doi.org/10.1103/PhysRevLett.121.161101}{\detokenize{10.1103/PhysRevLett.121.161101}}}.

\bibitem[Abbott \em{et~al.}(2017)Abbott, Abbott, Abbott, Acernese, Ackley,
  Adams, Adams, Addesso, Adhikari, Adya, Affeldt, Afrough, Agarwal, Agathos,
  Agatsuma, Aggarwal, Aguiar, Aiello, Ain, Ajith, Allen, Allen, Allocca, Altin,
  Amato, Ananyeva, Anderson, Anderson, Angelova, Antier, Appert, Arai, Araya,
  Areeda, Arnaud, Arun, Ascenzi, Ashton, Ast, Aston, Astone, Atallah, Aufmuth,
  Aulbert, AultONeal, Austin, Avila-Alvarez, Babak, Bacon, Bader, Bae, Bailes,
  Baker, Baldaccini, Ballardin, Ballmer, Banagiri, Barayoga, Barclay, Barish,
  Barker, Barkett, Barone, Barr, Barsotti, Barsuglia, Barta, Barthelmy,
  Bartlett, Bartos, Bassiri, Basti, Batch, Bawaj, Bayley, Bazzan, B\'ecsy,
  Beer, Bejger, Belahcene, Bell, Berger, Bergmann, Bernuzzi, Bero, Berry,
  Bersanetti, Bertolini, Betzwieser, Bhagwat, Bhandare, Bilenko, Billingsley,
  Billman, Birch, Birney, Birnholtz, Biscans, Biscoveanu, Bisht, Bitossi,
  Biwer, Bizouard, Blackburn, Blackman, Blair, Blair, Blair, Bloemen, Bock,
  Bode, Boer, Bogaert, Bohe, Bondu, Bonilla, Bonnand, Boom, Bork, Boschi, Bose,
  Bossie, Bouffanais, Bozzi, Bradaschia, Brady, Branchesi, Brau, Briant,
  Brillet, Brinkmann, Brisson, Brockill, Broida, Brooks, Brown, Brown, Brunett,
  Buchanan, Buikema, Bulik, Bulten, Buonanno, Buskulic, Buy, Byer, Cabero,
  Cadonati, Cagnoli, Cahillane, Calder\'on~Bustillo, Callister, Calloni, Camp,
  Canepa, Canizares, Cannon, Cao, Cao, Capano, Capocasa, Carbognani, Caride,
  Carney, Carullo, Casanueva~Diaz, Casentini, Caudill, Cavagli\`a, Cavalier,
  Cavalieri, Cella, Cepeda, Cerd\'a-Dur\'an, Cerretani, Cesarini, Chamberlin,
  Chan, Chao, Charlton, Chase, Chassande-Mottin, Chatterjee, Chatziioannou,
  Cheeseboro, Chen, Chen, Chen, Cheng, Chia, Chincarini, Chiummo, Chmiel, Cho,
  Cho, Chow, Christensen, Chu, Chua, Chua, Chung, Chung, Ciani, Ciolfi,
  Cirelli, Cirone, Clara, Clark, Clearwater, Cleva, Cocchieri, Coccia, Cohadon,
  Cohen, Colla, Collette, Cominsky, Constancio, Conti, Cooper, Corban, Corbitt,
  Cordero-Carri\'on, Corley, Cornish, Corsi, Cortese, Costa, Coughlin,
  Coughlin, Coulon, Countryman, Couvares, Covas, Cowan, Coward, Cowart, Coyne,
  Coyne, Creighton, Creighton, Cripe, Crowder, Cullen, Cumming, Cunningham,
  Cuoco, Dal~Canton, D\'alya, Danilishin, D'Antonio, Danzmann, Dasgupta,
  Da~Silva~Costa, Dattilo, Dave, Davier, Davis, Daw, Day, De, DeBra, Degallaix,
  De~Laurentis, Del\'eglise, Del~Pozzo, Demos, Denker, Dent, De~Pietri,
  Dergachev, De~Rosa, DeRosa, De~Rossi, DeSalvo, de~Varona, Devenson,
  Dhurandhar, D\'{\i}az, Dietrich, Di~Fiore, Di~Giovanni, Di~Girolamo,
  Di~Lieto, Di~Pace, Di~Palma, Di~Renzo, Doctor, Dolique, Donovan, Dooley,
  Doravari, Dorrington, Douglas, Dovale~\'Alvarez, Downes, Drago,
  Dreissigacker, Driggers, Du, Ducrot, Dudi, Dupej, Dwyer, Edo, Edwards,
  Effler, Eggenstein, Ehrens, Eichholz, Eikenberry, Eisenstein, Essick,
  Estevez, Etienne, Etzel, Evans, Evans, Factourovich, Fafone, Fair, Fairhurst,
  Fan, Farinon, Farr, Farr, Fauchon-Jones, Favata, Fays, Fee, Fehrmann, Feicht,
  Fejer, Fernandez-Galiana, Ferrante, Ferreira, Ferrini, Fidecaro, Finstad,
  Fiori, Fiorucci, Fishbach, Fisher, Fitz-Axen, Flaminio, Fletcher, Fong, Font,
  Forsyth, Forsyth, Fournier, Frasca, Frasconi, Frei, Freise, Frey, Frey,
  Fries, Fritschel, Frolov, Fulda, Fyffe, Gabbard, Gadre, Gaebel, Gair,
  Gammaitoni, Ganija, Gaonkar, Garcia-Quiros, Garufi, Gateley, Gaudio, Gaur,
  Gayathri, Gehrels, Gemme, Genin, Gennai, George, George, Gergely, Germain,
  Ghonge, Ghosh, Ghosh, Ghosh, Giaime, Giardina, Giazotto, Gill, Glover, Goetz,
  Goetz, Gomes, Goncharov, Gonz\'alez, Gonzalez~Castro, Gopakumar, Gorodetsky,
  Gossan, Gosselin, Gouaty, Grado, Graef, Granata, Grant, Gras, Gray, Greco,
  Green, Gretarsson, Groot, Grote, Grunewald, Gruning, Guidi, Guo, Gupta,
  Gupta, Gushwa, Gustafson, Gustafson, Halim, Hall, Hall, Hamilton, Hammond,
  Haney, Hanke, Hanks, Hanna, Hannam, Hannuksela, Hanson, Hardwick, Harms,
  Harry, Harry, Hart, Haster, Haughian, Healy, Heidmann, Heintze, Heitmann,
  Hello, Hemming, Hendry, Heng, Hennig, Heptonstall, Heurs, Hild, Hinderer, Ho,
  Hoak, Hofman, Holt, Holz, Hopkins, Horst, Hough, Houston, Howell, Hreibi, Hu,
  Huerta, Huet, Hughey, Husa, Huttner, Huynh-Dinh, Indik, Inta, Intini, Isa,
  Isac, Isi, Iyer, Izumi, Jacqmin, Jani, Jaranowski, Jawahar,
  Jim\'enez-Forteza, Johnson, Johnson-McDaniel, Jones, Jones, Jonker, Ju,
  Junker, Kalaghatgi, Kalogera, Kamai, Kandhasamy, Kang, Kanner, Kapadia,
  Karki, Karvinen, Kasprzack, Kastaun, Katolik, Katsavounidis, Katzman, Kaufer,
  Kawabe, K\'ef\'elian, Keitel, Kemball, Kennedy, Kent, Key, Khalili, Khan,
  Khan, Khan, Khazanov, Kijbunchoo, Kim, Kim, Kim, Kim, Kim, Kim, Kimbrell,
  King, King, Kinley-Hanlon, Kirchhoff, Kissel, Kleybolte, Klimenko, Knowles,
  Koch, Koehlenbeck, Koley, Kondrashov, Kontos, Korobko, Korth, Kowalska,
  Kozak, Kr\"amer, Kringel, Krishnan, Kr\'olak, Kuehn, Kumar, Kumar, Kumar,
  Kuo, Kutynia, Kwang, Lackey, Lai, Landry, Lang, Lange, Lantz, Lanza, Larson,
  Lartaux-Vollard, Lasky, Laxen, Lazzarini, Lazzaro, Leaci, Leavey, Lee, Lee,
  Lee, Lee, Lee, Lehmann, Lenon, Leon, Leonardi, Leroy, Letendre, Levin, Li,
  Linker, Littenberg, Liu, Liu, Lo, Lockerbie, London, Lord, Lorenzini,
  Loriette, Lormand, Losurdo, Lough, Lousto, Lovelace, L\"uck, Lumaca,
  Lundgren, Lynch, Ma, Macas, Macfoy, Machenschalk, MacInnis, Macleod, Maga\~na
  Hernandez, Maga\~na Sandoval, Maga\~na Zertuche, Magee, Majorana, Maksimovic,
  Man, Mandic, Mangano, Mansell, Manske, Mantovani, Marchesoni, Marion,
  M\'arka, M\'arka, Markakis, Markosyan, Markowitz, Maros, Marquina, Marsh,
  Martelli, Martellini, Martin, Martin, Martynov, Marx, Mason, Massera,
  Masserot, Massinger, Masso-Reid, Mastrogiovanni, Matas, Matichard, Matone,
  Mavalvala, Mazumder, McCarthy, McClelland, McCormick, McCuller, McGuire,
  McIntyre, McIver, McManus, McNeill, McRae, McWilliams, Meacher, Meadors,
  Mehmet, Meidam, Mejuto-Villa, Melatos, Mendell, Mercer, Merilh, Merzougui,
  Meshkov, Messenger, Messick, Metzdorff, Meyers, Miao, Michel, Middleton,
  Mikhailov, Milano, Miller, Miller, Miller, Millhouse, Milovich-Goff,
  Minazzoli, Minenkov, Ming, Mishra, Mitra, Mitrofanov, Mitselmakher,
  Mittleman, Moffa, Moggi, Mogushi, Mohan, Mohapatra, Molina, Montani, Moore,
  Moraru, Moreno, Morisaki, Morriss, Mours, Mow-Lowry, Mueller, Muir,
  Mukherjee, Mukherjee, Mukherjee, Mukund, Mullavey, Munch, Mu\~niz, Muratore,
  Murray, Nagar, Napier, Nardecchia, Naticchioni, Nayak, Neilson, Nelemans,
  Nelson, Nery, Neunzert, Nevin, Newport, Newton, Ng, Nguyen, Nguyen, Nichols,
  Nielsen, Nissanke, Nitz, Noack, Nocera, Nolting, North, Nuttall, Oberling,
  O'Dea, Ogin, Oh, Oh, Ohme, Okada, Oliver, Oppermann, Oram, O'Reilly,
  Ormiston, Ortega, O'Shaughnessy, Ossokine, Ottaway, Overmier, Owen, Pace,
  Page, Page, Pai, Pai, Palamos, Palashov, Palomba, Pal-Singh, Pan, Pan, Pang,
  Pang, Pankow, Pannarale, Pant, Paoletti, Paoli, Papa, Parida, Parker,
  Pascucci, Pasqualetti, Passaquieti, Passuello, Patil, Patricelli, Pearlstone,
  Pedraza, Pedurand, Pekowsky, Pele, Penn, Perez, Perreca, Perri, Pfeiffer,
  Phelps, Piccinni, Pichot, Piergiovanni, Pierro, Pillant, Pinard, Pinto,
  Pirello, Pitkin, Poe, Poggiani, Popolizio, Porter, Post, Powell, Prasad,
  Pratt, Pratten, Predoi, Prestegard, Prijatelj, Principe, Privitera, Prix,
  Prodi, Prokhorov, Puncken, Punturo, Puppo, P\"urrer, Qi, Quetschke, Quintero,
  Quitzow-James, Raab, Rabeling, Radkins, Raffai, Raja, Rajan, Rajbhandari,
  Rakhmanov, Ramirez, Ramos-Buades, Rapagnani, Raymond, Razzano, Read,
  Regimbau, Rei, Reid, Reitze, Ren, Reyes, Ricci, Ricker, Rieger, Riles, Rizzo,
  Robertson, Robie, Robinet, Rocchi, Rolland, Rollins, Roma, Romano, Romano,
  Romel, Romie, Rosi\ifmmode~\acute{n}\else \'{n}\fi{}ska, Ross, Rowan,
  R\"udiger, Ruggi, Rutins, Ryan, Sachdev, Sadecki, Sadeghian, Sakellariadou,
  Salconi, Saleem, Salemi, Samajdar, Sammut, Sampson, Sanchez, Sanchez,
  Sanchis-Gual, Sandberg, Sanders, Sassolas, Sathyaprakash, Saulson, Sauter,
  Savage, Sawadsky, Schale, Scheel, Scheuer, Schmidt, Schmidt, Schnabel,
  Schofield, Sch\"onbeck, Schreiber, Schuette, Schulte, Schutz, Schwalbe,
  Scott, Scott, Seidel, Sellers, Sengupta, Sentenac, Sequino, Sergeev,
  Shaddock, Shaffer, Shah, Shahriar, Shaner, Shao, Shapiro, Shawhan, Sheperd,
  Shoemaker, Shoemaker, Siellez, Siemens, Sieniawska, Sigg, Silva, Singer,
  Singh, Singhal, Sintes, Slagmolen, Smith, Smith, Smith, Somala, Son,
  Sonnenberg, Sorazu, Sorrentino, Souradeep, Spencer, Srivastava, Staats,
  Staley, Steinke, Steinlechner, Steinlechner, Steinmeyer, Stevenson, Stone,
  Stops, Strain, Stratta, Strigin, Strunk, Sturani, Stuver, Summerscales, Sun,
  Sunil, Suresh, Sutton, Swinkels, Szczepa\ifmmode~\acute{n}\else
  \'{n}\fi{}czyk, Tacca, Tait, Talbot, Talukder, Tanner, T\'apai, Taracchini,
  Tasson, Taylor, Taylor, Tewari, Theeg, Thies, Thomas, Thomas, Thomas, Thorne,
  Thorne, Thrane, Tiwari, Tiwari, Tokmakov, Toland, Tonelli, Tornasi,
  Torres-Forn\'e, Torrie, T\"oyr\"a, Travasso, Traylor, Trinastic, Tringali,
  Trozzo, Tsang, Tse, Tso, Tsukada, Tsuna, Tuyenbayev, Ueno, Ugolini,
  Unnikrishnan, Urban, Usman, Vahlbruch, Vajente, Valdes, Vallisneri, van
  Bakel, van Beuzekom, van~den Brand, Van Den~Broeck, Vander-Hyde, van~der
  Schaaf, van Heijningen, van Veggel, Vardaro, Varma, Vass, Vas\'uth, Vecchio,
  Vedovato, Veitch, Veitch, Venkateswara, Venugopalan, Verkindt, Vetrano,
  Vicer\'e, Viets, Vinciguerra, Vine, Vinet, Vitale, Vo, Vocca, Vorvick,
  Vyatchanin, Wade, Wade, Wade, Walet, Walker, Wallace, Walsh, Wang, Wang,
  Wang, Wang, Wang, Ward, Warner, Was, Watchi, Weaver, Wei, Weinert, Weinstein,
  Weiss, Wen, Wessel, We\ss{}els, Westerweck, Westphal, Wette, Whelan,
  Whitcomb, Whiting, Whittle, Wilken, Williams, Williams, Williamson, Willis,
  Willke, Wimmer, Winkler, Wipf, Wittel, Woan, Woehler, Wofford, Wong, Worden,
  Wright, Wu, Wysocki, Xiao, Yamamoto, Yancey, Yang, Yap, Yazback, Yu, Yu,
  Yvert, Zadro\ifmmode~\dot{z}\else \.{z}\fi{}ny, Zanolin, Zelenova, Zendri,
  Zevin, Zhang, Zhang, Zhang, Zhang, Zhao, Zhou, Zhou, Zhu, Zhu, Zimmerman,
  Zucker, and Zweizig]{LIGO_NSNSGW_detection}
Abbott, B.P.; Abbott, R.; Abbott, T.D.; Acernese, F.; Ackley, K.; Adams, C.;
  Adams, T.; Addesso, P.; Adhikari, R.X.; Adya, V.B.; Affeldt, C.; Afrough, M.;
  Agarwal, B.; Agathos, M.; Agatsuma, K.; Aggarwal, N.; Aguiar, O.D.; Aiello,
  L.; Ain, A.; Ajith, P.; Allen, B.; Allen, G.; Allocca, A.; Altin, P.A.;
  Amato, A.; Ananyeva, A.; Anderson, S.B.; Anderson, W.G.; Angelova, S.V.;
  Antier, S.; Appert, S.; Arai, K.; Araya, M.C.; Areeda, J.S.; Arnaud, N.;
  Arun, K.G.; Ascenzi, S.; Ashton, G.; Ast, M.; Aston, S.M.; Astone, P.;
  Atallah, D.V.; Aufmuth, P.; Aulbert, C.; AultONeal, K.; Austin, C.;
  Avila-Alvarez, A.; Babak, S.; Bacon, P.; Bader, M.K.M.; Bae, S.; Bailes, M.;
  Baker, P.T.; Baldaccini, F.; Ballardin, G.; Ballmer, S.W.; Banagiri, S.;
  Barayoga, J.C.; Barclay, S.E.; Barish, B.C.; Barker, D.; Barkett, K.; Barone,
  F.; Barr, B.; Barsotti, L.; Barsuglia, M.; Barta, D.; Barthelmy, S.D.;
  Bartlett, J.; Bartos, I.; Bassiri, R.; Basti, A.; Batch, J.C.; Bawaj, M.;
  Bayley, J.C.; Bazzan, M.; B\'ecsy, B.; Beer, C.; Bejger, M.; Belahcene, I.;
  Bell, A.S.; Berger, B.K.; Bergmann, G.; Bernuzzi, S.; Bero, J.J.; Berry,
  C.P.L.; Bersanetti, D.; Bertolini, A.; Betzwieser, J.; Bhagwat, S.; Bhandare,
  R.; Bilenko, I.A.; Billingsley, G.; Billman, C.R.; Birch, J.; Birney, R.;
  Birnholtz, O.; Biscans, S.; Biscoveanu, S.; Bisht, A.; Bitossi, M.; Biwer,
  C.; Bizouard, M.A.; Blackburn, J.K.; Blackman, J.; Blair, C.D.; Blair, D.G.;
  Blair, R.M.; Bloemen, S.; Bock, O.; Bode, N.; Boer, M.; Bogaert, G.; Bohe,
  A.; Bondu, F.; Bonilla, E.; Bonnand, R.; Boom, B.A.; Bork, R.; Boschi, V.;
  Bose, S.; Bossie, K.; Bouffanais, Y.; Bozzi, A.; Bradaschia, C.; Brady, P.R.;
  Branchesi, M.; Brau, J.E.; Briant, T.; Brillet, A.; Brinkmann, M.; Brisson,
  V.; Brockill, P.; Broida, J.E.; Brooks, A.F.; Brown, D.A.; Brown, D.D.;
  Brunett, S.; Buchanan, C.C.; Buikema, A.; Bulik, T.; Bulten, H.J.; Buonanno,
  A.; Buskulic, D.; Buy, C.; Byer, R.L.; Cabero, M.; Cadonati, L.; Cagnoli, G.;
  Cahillane, C.; Calder\'on~Bustillo, J.; Callister, T.A.; Calloni, E.; Camp,
  J.B.; Canepa, M.; Canizares, P.; Cannon, K.C.; Cao, H.; Cao, J.; Capano,
  C.D.; Capocasa, E.; Carbognani, F.; Caride, S.; Carney, M.F.; Carullo, G.;
  Casanueva~Diaz, J.; Casentini, C.; Caudill, S.; Cavagli\`a, M.; Cavalier, F.;
  Cavalieri, R.; Cella, G.; Cepeda, C.B.; Cerd\'a-Dur\'an, P.; Cerretani, G.;
  Cesarini, E.; Chamberlin, S.J.; Chan, M.; Chao, S.; Charlton, P.; Chase, E.;
  Chassande-Mottin, E.; Chatterjee, D.; Chatziioannou, K.; Cheeseboro, B.D.;
  Chen, H.Y.; Chen, X.; Chen, Y.; Cheng, H.P.; Chia, H.; Chincarini, A.;
  Chiummo, A.; Chmiel, T.; Cho, H.S.; Cho, M.; Chow, J.H.; Christensen, N.;
  Chu, Q.; Chua, A.J.K.; Chua, S.; Chung, A.K.W.; Chung, S.; Ciani, G.; Ciolfi,
  R.; Cirelli, C.E.; Cirone, A.; Clara, F.; Clark, J.A.; Clearwater, P.; Cleva,
  F.; Cocchieri, C.; Coccia, E.; Cohadon, P.F.; Cohen, D.; Colla, A.; Collette,
  C.G.; Cominsky, L.R.; Constancio, M.; Conti, L.; Cooper, S.J.; Corban, P.;
  Corbitt, T.R.; Cordero-Carri\'on, I.; Corley, K.R.; Cornish, N.; Corsi, A.;
  Cortese, S.; Costa, C.A.; Coughlin, M.W.; Coughlin, S.B.; Coulon, J.P.;
  Countryman, S.T.; Couvares, P.; Covas, P.B.; Cowan, E.E.; Coward, D.M.;
  Cowart, M.J.; Coyne, D.C.; Coyne, R.; Creighton, J.D.E.; Creighton, T.D.;
  Cripe, J.; Crowder, S.G.; Cullen, T.J.; Cumming, A.; Cunningham, L.; Cuoco,
  E.; Dal~Canton, T.; D\'alya, G.; Danilishin, S.L.; D'Antonio, S.; Danzmann,
  K.; Dasgupta, A.; Da~Silva~Costa, C.F.; Dattilo, V.; Dave, I.; Davier, M.;
  Davis, D.; Daw, E.J.; Day, B.; De, S.; DeBra, D.; Degallaix, J.;
  De~Laurentis, M.; Del\'eglise, S.; Del~Pozzo, W.; Demos, N.; Denker, T.;
  Dent, T.; De~Pietri, R.; Dergachev, V.; De~Rosa, R.; DeRosa, R.T.; De~Rossi,
  C.; DeSalvo, R.; de~Varona, O.; Devenson, J.; Dhurandhar, S.; D\'{\i}az,
  M.C.; Dietrich, T.; Di~Fiore, L.; Di~Giovanni, M.; Di~Girolamo, T.; Di~Lieto,
  A.; Di~Pace, S.; Di~Palma, I.; Di~Renzo, F.; Doctor, Z.; Dolique, V.;
  Donovan, F.; Dooley, K.L.; Doravari, S.; Dorrington, I.; Douglas, R.;
  Dovale~\'Alvarez, M.; Downes, T.P.; Drago, M.; Dreissigacker, C.; Driggers,
  J.C.; Du, Z.; Ducrot, M.; Dudi, R.; Dupej, P.; Dwyer, S.E.; Edo, T.B.;
  Edwards, M.C.; Effler, A.; Eggenstein, H.B.; Ehrens, P.; Eichholz, J.;
  Eikenberry, S.S.; Eisenstein, R.A.; Essick, R.C.; Estevez, D.; Etienne, Z.B.;
  Etzel, T.; Evans, M.; Evans, T.M.; Factourovich, M.; Fafone, V.; Fair, H.;
  Fairhurst, S.; Fan, X.; Farinon, S.; Farr, B.; Farr, W.M.; Fauchon-Jones,
  E.J.; Favata, M.; Fays, M.; Fee, C.; Fehrmann, H.; Feicht, J.; Fejer, M.M.;
  Fernandez-Galiana, A.; Ferrante, I.; Ferreira, E.C.; Ferrini, F.; Fidecaro,
  F.; Finstad, D.; Fiori, I.; Fiorucci, D.; Fishbach, M.; Fisher, R.P.;
  Fitz-Axen, M.; Flaminio, R.; Fletcher, M.; Fong, H.; Font, J.A.; Forsyth,
  P.W.F.; Forsyth, S.S.; Fournier, J.D.; Frasca, S.; Frasconi, F.; Frei, Z.;
  Freise, A.; Frey, R.; Frey, V.; Fries, E.M.; Fritschel, P.; Frolov, V.V.;
  Fulda, P.; Fyffe, M.; Gabbard, H.; Gadre, B.U.; Gaebel, S.M.; Gair, J.R.;
  Gammaitoni, L.; Ganija, M.R.; Gaonkar, S.G.; Garcia-Quiros, C.; Garufi, F.;
  Gateley, B.; Gaudio, S.; Gaur, G.; Gayathri, V.; Gehrels, N.; Gemme, G.;
  Genin, E.; Gennai, A.; George, D.; George, J.; Gergely, L.; Germain, V.;
  Ghonge, S.; Ghosh, A.; Ghosh, A.; Ghosh, S.; Giaime, J.A.; Giardina, K.D.;
  Giazotto, A.; Gill, K.; Glover, L.; Goetz, E.; Goetz, R.; Gomes, S.;
  Goncharov, B.; Gonz\'alez, G.; Gonzalez~Castro, J.M.; Gopakumar, A.;
  Gorodetsky, M.L.; Gossan, S.E.; Gosselin, M.; Gouaty, R.; Grado, A.; Graef,
  C.; Granata, M.; Grant, A.; Gras, S.; Gray, C.; Greco, G.; Green, A.C.;
  Gretarsson, E.M.; Groot, P.; Grote, H.; Grunewald, S.; Gruning, P.; Guidi,
  G.M.; Guo, X.; Gupta, A.; Gupta, M.K.; Gushwa, K.E.; Gustafson, E.K.;
  Gustafson, R.; Halim, O.; Hall, B.R.; Hall, E.D.; Hamilton, E.Z.; Hammond,
  G.; Haney, M.; Hanke, M.M.; Hanks, J.; Hanna, C.; Hannam, M.D.; Hannuksela,
  O.A.; Hanson, J.; Hardwick, T.; Harms, J.; Harry, G.M.; Harry, I.W.; Hart,
  M.J.; Haster, C.J.; Haughian, K.; Healy, J.; Heidmann, A.; Heintze, M.C.;
  Heitmann, H.; Hello, P.; Hemming, G.; Hendry, M.; Heng, I.S.; Hennig, J.;
  Heptonstall, A.W.; Heurs, M.; Hild, S.; Hinderer, T.; Ho, W.C.G.; Hoak, D.;
  Hofman, D.; Holt, K.; Holz, D.E.; Hopkins, P.; Horst, C.; Hough, J.; Houston,
  E.A.; Howell, E.J.; Hreibi, A.; Hu, Y.M.; Huerta, E.A.; Huet, D.; Hughey, B.;
  Husa, S.; Huttner, S.H.; Huynh-Dinh, T.; Indik, N.; Inta, R.; Intini, G.;
  Isa, H.N.; Isac, J.M.; Isi, M.; Iyer, B.R.; Izumi, K.; Jacqmin, T.; Jani, K.;
  Jaranowski, P.; Jawahar, S.; Jim\'enez-Forteza, F.; Johnson, W.W.;
  Johnson-McDaniel, N.K.; Jones, D.I.; Jones, R.; Jonker, R.J.G.; Ju, L.;
  Junker, J.; Kalaghatgi, C.V.; Kalogera, V.; Kamai, B.; Kandhasamy, S.; Kang,
  G.; Kanner, J.B.; Kapadia, S.J.; Karki, S.; Karvinen, K.S.; Kasprzack, M.;
  Kastaun, W.; Katolik, M.; Katsavounidis, E.; Katzman, W.; Kaufer, S.; Kawabe,
  K.; K\'ef\'elian, F.; Keitel, D.; Kemball, A.J.; Kennedy, R.; Kent, C.; Key,
  J.S.; Khalili, F.Y.; Khan, I.; Khan, S.; Khan, Z.; Khazanov, E.A.;
  Kijbunchoo, N.; Kim, C.; Kim, J.C.; Kim, K.; Kim, W.; Kim, W.S.; Kim, Y.M.;
  Kimbrell, S.J.; King, E.J.; King, P.J.; Kinley-Hanlon, M.; Kirchhoff, R.;
  Kissel, J.S.; Kleybolte, L.; Klimenko, S.; Knowles, T.D.; Koch, P.;
  Koehlenbeck, S.M.; Koley, S.; Kondrashov, V.; Kontos, A.; Korobko, M.; Korth,
  W.Z.; Kowalska, I.; Kozak, D.B.; Kr\"amer, C.; Kringel, V.; Krishnan, B.;
  Kr\'olak, A.; Kuehn, G.; Kumar, P.; Kumar, R.; Kumar, S.; Kuo, L.; Kutynia,
  A.; Kwang, S.; Lackey, B.D.; Lai, K.H.; Landry, M.; Lang, R.N.; Lange, J.;
  Lantz, B.; Lanza, R.K.; Larson, S.L.; Lartaux-Vollard, A.; Lasky, P.D.;
  Laxen, M.; Lazzarini, A.; Lazzaro, C.; Leaci, P.; Leavey, S.; Lee, C.H.; Lee,
  H.K.; Lee, H.M.; Lee, H.W.; Lee, K.; Lehmann, J.; Lenon, A.; Leon, E.;
  Leonardi, M.; Leroy, N.; Letendre, N.; Levin, Y.; Li, T.G.F.; Linker, S.D.;
  Littenberg, T.B.; Liu, J.; Liu, X.; Lo, R.K.L.; Lockerbie, N.A.; London,
  L.T.; Lord, J.E.; Lorenzini, M.; Loriette, V.; Lormand, M.; Losurdo, G.;
  Lough, J.D.; Lousto, C.O.; Lovelace, G.; L\"uck, H.; Lumaca, D.; Lundgren,
  A.P.; Lynch, R.; Ma, Y.; Macas, R.; Macfoy, S.; Machenschalk, B.; MacInnis,
  M.; Macleod, D.M.; Maga\~na Hernandez, I.; Maga\~na Sandoval, F.; Maga\~na
  Zertuche, L.; Magee, R.M.; Majorana, E.; Maksimovic, I.; Man, N.; Mandic, V.;
  Mangano, V.; Mansell, G.L.; Manske, M.; Mantovani, M.; Marchesoni, F.;
  Marion, F.; M\'arka, S.; M\'arka, Z.; Markakis, C.; Markosyan, A.S.;
  Markowitz, A.; Maros, E.; Marquina, A.; Marsh, P.; Martelli, F.; Martellini,
  L.; Martin, I.W.; Martin, R.M.; Martynov, D.V.; Marx, J.N.; Mason, K.;
  Massera, E.; Masserot, A.; Massinger, T.J.; Masso-Reid, M.; Mastrogiovanni,
  S.; Matas, A.; Matichard, F.; Matone, L.; Mavalvala, N.; Mazumder, N.;
  McCarthy, R.; McClelland, D.E.; McCormick, S.; McCuller, L.; McGuire, S.C.;
  McIntyre, G.; McIver, J.; McManus, D.J.; McNeill, L.; McRae, T.; McWilliams,
  S.T.; Meacher, D.; Meadors, G.D.; Mehmet, M.; Meidam, J.; Mejuto-Villa, E.;
  Melatos, A.; Mendell, G.; Mercer, R.A.; Merilh, E.L.; Merzougui, M.; Meshkov,
  S.; Messenger, C.; Messick, C.; Metzdorff, R.; Meyers, P.M.; Miao, H.;
  Michel, C.; Middleton, H.; Mikhailov, E.E.; Milano, L.; Miller, A.L.; Miller,
  B.B.; Miller, J.; Millhouse, M.; Milovich-Goff, M.C.; Minazzoli, O.;
  Minenkov, Y.; Ming, J.; Mishra, C.; Mitra, S.; Mitrofanov, V.P.;
  Mitselmakher, G.; Mittleman, R.; Moffa, D.; Moggi, A.; Mogushi, K.; Mohan,
  M.; Mohapatra, S.R.P.; Molina, I.; Montani, M.; Moore, C.J.; Moraru, D.;
  Moreno, G.; Morisaki, S.; Morriss, S.R.; Mours, B.; Mow-Lowry, C.M.; Mueller,
  G.; Muir, A.W.; Mukherjee, A.; Mukherjee, D.; Mukherjee, S.; Mukund, N.;
  Mullavey, A.; Munch, J.; Mu\~niz, E.A.; Muratore, M.; Murray, P.G.; Nagar,
  A.; Napier, K.; Nardecchia, I.; Naticchioni, L.; Nayak, R.K.; Neilson, J.;
  Nelemans, G.; Nelson, T.J.N.; Nery, M.; Neunzert, A.; Nevin, L.; Newport,
  J.M.; Newton, G.; Ng, K.K.Y.; Nguyen, P.; Nguyen, T.T.; Nichols, D.; Nielsen,
  A.B.; Nissanke, S.; Nitz, A.; Noack, A.; Nocera, F.; Nolting, D.; North, C.;
  Nuttall, L.K.; Oberling, J.; O'Dea, G.D.; Ogin, G.H.; Oh, J.J.; Oh, S.H.;
  Ohme, F.; Okada, M.A.; Oliver, M.; Oppermann, P.; Oram, R.J.; O'Reilly, B.;
  Ormiston, R.; Ortega, L.F.; O'Shaughnessy, R.; Ossokine, S.; Ottaway, D.J.;
  Overmier, H.; Owen, B.J.; Pace, A.E.; Page, J.; Page, M.A.; Pai, A.; Pai,
  S.A.; Palamos, J.R.; Palashov, O.; Palomba, C.; Pal-Singh, A.; Pan, H.; Pan,
  H.W.; Pang, B.; Pang, P.T.H.; Pankow, C.; Pannarale, F.; Pant, B.C.;
  Paoletti, F.; Paoli, A.; Papa, M.A.; Parida, A.; Parker, W.; Pascucci, D.;
  Pasqualetti, A.; Passaquieti, R.; Passuello, D.; Patil, M.; Patricelli, B.;
  Pearlstone, B.L.; Pedraza, M.; Pedurand, R.; Pekowsky, L.; Pele, A.; Penn,
  S.; Perez, C.J.; Perreca, A.; Perri, L.M.; Pfeiffer, H.P.; Phelps, M.;
  Piccinni, O.J.; Pichot, M.; Piergiovanni, F.; Pierro, V.; Pillant, G.;
  Pinard, L.; Pinto, I.M.; Pirello, M.; Pitkin, M.; Poe, M.; Poggiani, R.;
  Popolizio, P.; Porter, E.K.; Post, A.; Powell, J.; Prasad, J.; Pratt, J.W.W.;
  Pratten, G.; Predoi, V.; Prestegard, T.; Prijatelj, M.; Principe, M.;
  Privitera, S.; Prix, R.; Prodi, G.A.; Prokhorov, L.G.; Puncken, O.; Punturo,
  M.; Puppo, P.; P\"urrer, M.; Qi, H.; Quetschke, V.; Quintero, E.A.;
  Quitzow-James, R.; Raab, F.J.; Rabeling, D.S.; Radkins, H.; Raffai, P.; Raja,
  S.; Rajan, C.; Rajbhandari, B.; Rakhmanov, M.; Ramirez, K.E.; Ramos-Buades,
  A.; Rapagnani, P.; Raymond, V.; Razzano, M.; Read, J.; Regimbau, T.; Rei, L.;
  Reid, S.; Reitze, D.H.; Ren, W.; Reyes, S.D.; Ricci, F.; Ricker, P.M.;
  Rieger, S.; Riles, K.; Rizzo, M.; Robertson, N.A.; Robie, R.; Robinet, F.;
  Rocchi, A.; Rolland, L.; Rollins, J.G.; Roma, V.J.; Romano, J.D.; Romano, R.;
  Romel, C.L.; Romie, J.H.; Rosi\ifmmode~\acute{n}\else \'{n}\fi{}ska, D.;
  Ross, M.P.; Rowan, S.; R\"udiger, A.; Ruggi, P.; Rutins, G.; Ryan, K.;
  Sachdev, S.; Sadecki, T.; Sadeghian, L.; Sakellariadou, M.; Salconi, L.;
  Saleem, M.; Salemi, F.; Samajdar, A.; Sammut, L.; Sampson, L.M.; Sanchez,
  E.J.; Sanchez, L.E.; Sanchis-Gual, N.; Sandberg, V.; Sanders, J.R.; Sassolas,
  B.; Sathyaprakash, B.S.; Saulson, P.R.; Sauter, O.; Savage, R.L.; Sawadsky,
  A.; Schale, P.; Scheel, M.; Scheuer, J.; Schmidt, J.; Schmidt, P.; Schnabel,
  R.; Schofield, R.M.S.; Sch\"onbeck, A.; Schreiber, E.; Schuette, D.; Schulte,
  B.W.; Schutz, B.F.; Schwalbe, S.G.; Scott, J.; Scott, S.M.; Seidel, E.;
  Sellers, D.; Sengupta, A.S.; Sentenac, D.; Sequino, V.; Sergeev, A.;
  Shaddock, D.A.; Shaffer, T.J.; Shah, A.A.; Shahriar, M.S.; Shaner, M.B.;
  Shao, L.; Shapiro, B.; Shawhan, P.; Sheperd, A.; Shoemaker, D.H.; Shoemaker,
  D.M.; Siellez, K.; Siemens, X.; Sieniawska, M.; Sigg, D.; Silva, A.D.;
  Singer, L.P.; Singh, A.; Singhal, A.; Sintes, A.M.; Slagmolen, B.J.J.; Smith,
  B.; Smith, J.R.; Smith, R.J.E.; Somala, S.; Son, E.J.; Sonnenberg, J.A.;
  Sorazu, B.; Sorrentino, F.; Souradeep, T.; Spencer, A.P.; Srivastava, A.K.;
  Staats, K.; Staley, A.; Steinke, M.; Steinlechner, J.; Steinlechner, S.;
  Steinmeyer, D.; Stevenson, S.P.; Stone, R.; Stops, D.J.; Strain, K.A.;
  Stratta, G.; Strigin, S.E.; Strunk, A.; Sturani, R.; Stuver, A.L.;
  Summerscales, T.Z.; Sun, L.; Sunil, S.; Suresh, J.; Sutton, P.J.; Swinkels,
  B.L.; Szczepa\ifmmode~\acute{n}\else \'{n}\fi{}czyk, M.J.; Tacca, M.; Tait,
  S.C.; Talbot, C.; Talukder, D.; Tanner, D.B.; T\'apai, M.; Taracchini, A.;
  Tasson, J.D.; Taylor, J.A.; Taylor, R.; Tewari, S.V.; Theeg, T.; Thies, F.;
  Thomas, E.G.; Thomas, M.; Thomas, P.; Thorne, K.A.; Thorne, K.S.; Thrane, E.;
  Tiwari, S.; Tiwari, V.; Tokmakov, K.V.; Toland, K.; Tonelli, M.; Tornasi, Z.;
  Torres-Forn\'e, A.; Torrie, C.I.; T\"oyr\"a, D.; Travasso, F.; Traylor, G.;
  Trinastic, J.; Tringali, M.C.; Trozzo, L.; Tsang, K.W.; Tse, M.; Tso, R.;
  Tsukada, L.; Tsuna, D.; Tuyenbayev, D.; Ueno, K.; Ugolini, D.; Unnikrishnan,
  C.S.; Urban, A.L.; Usman, S.A.; Vahlbruch, H.; Vajente, G.; Valdes, G.;
  Vallisneri, M.; van Bakel, N.; van Beuzekom, M.; van~den Brand, J.F.J.; Van
  Den~Broeck, C.; Vander-Hyde, D.C.; van~der Schaaf, L.; van Heijningen, J.V.;
  van Veggel, A.A.; Vardaro, M.; Varma, V.; Vass, S.; Vas\'uth, M.; Vecchio,
  A.; Vedovato, G.; Veitch, J.; Veitch, P.J.; Venkateswara, K.; Venugopalan,
  G.; Verkindt, D.; Vetrano, F.; Vicer\'e, A.; Viets, A.D.; Vinciguerra, S.;
  Vine, D.J.; Vinet, J.Y.; Vitale, S.; Vo, T.; Vocca, H.; Vorvick, C.;
  Vyatchanin, S.P.; Wade, A.R.; Wade, L.E.; Wade, M.; Walet, R.; Walker, M.;
  Wallace, L.; Walsh, S.; Wang, G.; Wang, H.; Wang, J.Z.; Wang, W.H.; Wang,
  Y.F.; Ward, R.L.; Warner, J.; Was, M.; Watchi, J.; Weaver, B.; Wei, L.W.;
  Weinert, M.; Weinstein, A.J.; Weiss, R.; Wen, L.; Wessel, E.K.; We\ss{}els,
  P.; Westerweck, J.; Westphal, T.; Wette, K.; Whelan, J.T.; Whitcomb, S.E.;
  Whiting, B.F.; Whittle, C.; Wilken, D.; Williams, D.; Williams, R.D.;
  Williamson, A.R.; Willis, J.L.; Willke, B.; Wimmer, M.H.; Winkler, W.; Wipf,
  C.C.; Wittel, H.; Woan, G.; Woehler, J.; Wofford, J.; Wong, K.W.K.; Worden,
  J.; Wright, J.L.; Wu, D.S.; Wysocki, D.M.; Xiao, S.; Yamamoto, H.; Yancey,
  C.C.; Yang, L.; Yap, M.J.; Yazback, M.; Yu, H.; Yu, H.; Yvert, M.;
  Zadro\ifmmode~\dot{z}\else \.{z}\fi{}ny, A.; Zanolin, M.; Zelenova, T.;
  Zendri, J.P.; Zevin, M.; Zhang, L.; Zhang, M.; Zhang, T.; Zhang, Y.H.; Zhao,
  C.; Zhou, M.; Zhou, Z.; Zhu, S.J.; Zhu, X.J.; Zimmerman, A.B.; Zucker, M.E.;
  Zweizig, J.
\newblock GW170817: Observation of Gravitational Waves from a Binary Neutron
  Star Inspiral.
\newblock {\em Phys. Rev. Lett.} {\bf 2017}, {\em 119},~161101.
\newblock
  doi:{\changeurlcolor{black}\href{https://doi.org/10.1103/PhysRevLett.119.161101}{\detokenize{10.1103/PhysRevLett.119.161101}}}.

\bibitem[Rezzolla and Takami(2016)]{Rezzolla:2016nxn}
Rezzolla, L.; Takami, K.
\newblock {Gravitational-wave signal from binary neutron stars: a systematic
  analysis of the spectral properties}.
\newblock {\em Phys. Rev.} {\bf 2016}, {\em D93},~124051,
  \href{http://xxx.lanl.gov/abs/1604.00246}{{\normalfont
  [arXiv:gr-qc/1604.00246]}}.
\newblock
  doi:{\changeurlcolor{black}\href{https://doi.org/10.1103/PhysRevD.93.124051}{\detokenize{10.1103/PhysRevD.93.124051}}}.

\bibitem[Ozel \em{et~al.}(2016)Ozel, Psaltis, Guver, Baym, Heinke, and
  Guillot]{Ozel:2016}
Ozel, F.; Psaltis, D.; Guver, T.; Baym, G.; Heinke, C.; Guillot, S.
\newblock {The Dense Matter Equation of State from Neutron Star Radius and Mass
  Measurements}.
\newblock {\em Astrophys. J.} {\bf 2016}, {\em 820},~28,
  \href{http://xxx.lanl.gov/abs/1505.05155}{{\normalfont
  [arXiv:astro-ph.HE/1505.05155]}}.
\newblock
  doi:{\changeurlcolor{black}\href{https://doi.org/10.3847/0004-637X/820/1/28}{\detokenize{10.3847/0004-637X/820/1/28}}}.

\bibitem[Raithel \em{et~al.}(2017)Raithel, Özel, and Psaltis]{Raithel:2017ity}
Raithel, C.A.; Özel, F.; Psaltis, D.
\newblock {From Neutron Star Observables to the Equation of State. II. Bayesian
  Inference of Equation of State Pressures}.
\newblock {\em Astrophys. J.} {\bf 2017}, {\em 844},~156,
  \href{http://xxx.lanl.gov/abs/1704.00737}{{\normalfont
  [arXiv:astro-ph.HE/1704.00737]}}.
\newblock
  doi:{\changeurlcolor{black}\href{https://doi.org/10.3847/1538-4357/aa7a5a}{\detokenize{10.3847/1538-4357/aa7a5a}}}.

\bibitem[Guenther \em{et~al.}(2017)Guenther, Bellwied, Borsányi, Fodor, Katz,
  Pásztor, Ratti, and Szabó]{Katz_finite_mu_analitic_cont}
Guenther, J.; Bellwied, R.; Borsányi, S.; Fodor, Z.; Katz, S.; Pásztor, A.;
  Ratti, C.; Szabó, K.
\newblock The QCD equation of state at finite density from analytical
  continuation.
\newblock {\em Nuclear Physics A} {\bf 2017}, {\em 967},~720 -- 723.
\newblock The 26th International Conference on Ultra-relativistic
  Nucleus-Nucleus Collisions: Quark Matter 2017,
  doi:{\changeurlcolor{black}\href{https://doi.org/https://doi.org/10.1016/j.nuclphysa.2017.05.044}{\detokenize{https://doi.org/10.1016/j.nuclphysa.2017.05.044}}}.

\bibitem[{G\"unther, Jana} \em{et~al.}(2017){G\"unther, Jana}, {Bellwied,
  Rene}, {Borsanyi, Szabolcs}, {Fodor, Zoltan}, {Katz, Sandor D.}, {Pasztor,
  Attila}, and {Ratti, Claudia}]{katz_finite_mu_EoS}
{G\"unther, Jana}.; {Bellwied, Rene}.; {Borsanyi, Szabolcs}.; {Fodor, Zoltan}.;
  {Katz, Sandor D.}.; {Pasztor, Attila}.; {Ratti, Claudia}.
\newblock The QCD equation of state at finite density from analytical
  continuation.
\newblock {\em EPJ Web Conf.} {\bf 2017}, {\em 137},~07008.
\newblock
  doi:{\changeurlcolor{black}\href{https://doi.org/10.1051/epjconf/201713707008}{\detokenize{10.1051/epjconf/201713707008}}}.

\bibitem[Endr\ifmmode~\mbox{\H{o}}\else \H{o}\fi{}di
  \em{et~al.}(2018)Endr\ifmmode~\mbox{\H{o}}\else \H{o}\fi{}di, Fodor, Katz,
  Sexty, Szab\'o, and T\"or\"ok]{Katz_finite_mu_lattice}
Endr\ifmmode~\mbox{\H{o}}\else \H{o}\fi{}di, G.; Fodor, Z.; Katz, S.D.; Sexty,
  D.; Szab\'o, K.K.; T\"or\"ok, C.
\newblock Applying constrained simulations for low temperature lattice QCD at
  finite baryon chemical potential.
\newblock {\em Phys. Rev. D} {\bf 2018}, {\em 98},~074508.
\newblock
  doi:{\changeurlcolor{black}\href{https://doi.org/10.1103/PhysRevD.98.074508}{\detokenize{10.1103/PhysRevD.98.074508}}}.

\bibitem[Holt \em{et~al.}(2016)Holt, Rho, and Weise]{Holt:2014hma}
Holt, J.W.; Rho, M.; Weise, W.
\newblock {Chiral symmetry and effective field theories for hadronic, nuclear
  and stellar matter}.
\newblock {\em Phys. Rept.} {\bf 2016}, {\em 621},~2--75,
  \href{http://xxx.lanl.gov/abs/1411.6681}{{\normalfont
  [arXiv:nucl-th/1411.6681]}}.
\newblock
  doi:{\changeurlcolor{black}\href{https://doi.org/10.1016/j.physrep.2015.10.011}{\detokenize{10.1016/j.physrep.2015.10.011}}}.

\bibitem[Kojo(2017)]{Kojo:2017pfw}
Kojo, T.
\newblock {QCD in stars}.
\newblock  {10th International Workshop on Critical Point and Onset of
  Deconfinement (CPOD 2016) Wrocław, Poland, May 30-June 4, 2016},  2017,
  \href{http://xxx.lanl.gov/abs/1702.07997}{{\normalfont
  [arXiv:astro-ph.HE/1702.07997]}}.

\bibitem[Barnafoldi \em{et~al.}(2017)Barnafoldi, Jakovac, and
  Posfay]{harmonic:2017}
Barnafoldi, G.G.; Jakovac, A.; Posfay, P.
\newblock {Harmonic expansion of the effective potential in a functional
  renormalization group at finite chemical potential}.
\newblock {\em Phys. Rev.} {\bf 2017}, {\em D95},~025004,
  \href{http://xxx.lanl.gov/abs/1604.01717}{{\normalfont
  [arXiv:hep-th/1604.01717]}}.
\newblock
  doi:{\changeurlcolor{black}\href{https://doi.org/10.1103/PhysRevD.95.025004}{\detokenize{10.1103/PhysRevD.95.025004}}}.

\bibitem[Kov\'acs and Sz\'ep(2007)]{Zsolt:2007}
Kov\'acs, P.; Sz\'ep, Z.
\newblock Critical surface of the
  $SU(3{)}_{L}\ifmmode\times\else\texttimes\fi{}SU(3{)}_{R}$ chiral quark model
  at nonzero baryon density.
\newblock {\em Phys. Rev. D} {\bf 2007}, {\em 75},~025015.
\newblock
  doi:{\changeurlcolor{black}\href{https://doi.org/10.1103/PhysRevD.75.025015}{\detokenize{10.1103/PhysRevD.75.025015}}}.

\bibitem[Pósfay \em{et~al.}(2018{\natexlab{a}})Pósfay, Barnaföldi, and
  Jakovác]{Posfay:2017cor}
Pósfay, P.; Barnaföldi, G.G.; Jakovác, A.
\newblock {The effect of quantum fluctuations in compact star observables}.
\newblock {\em Publ. Astron. Soc. Austral.} {\bf 2018}, {\em 35},~19,
  \href{http://xxx.lanl.gov/abs/1710.05410}{{\normalfont
  [arXiv:hep-ph/1710.05410]}}.
\newblock
  doi:{\changeurlcolor{black}\href{https://doi.org/10.1017/pasa.2018.14}{\detokenize{10.1017/pasa.2018.14}}}.

\bibitem[Pósfay \em{et~al.}(2018{\natexlab{b}})Pósfay, Barnaföldi, and
  Jakovác]{Posfay:2016ygf}
Pósfay, P.; Barnaföldi, G.G.; Jakovác, A.
\newblock {Effect of quantum fluctuations in the high-energy cold nuclear
  equation of state and in compact star observables}.
\newblock {\em Phys. Rev.} {\bf 2018}, {\em C97},~025803,
  \href{http://xxx.lanl.gov/abs/1610.03674}{{\normalfont
  [arXiv:nucl-th/1610.03674]}}.
\newblock
  doi:{\changeurlcolor{black}\href{https://doi.org/10.1103/PhysRevC.97.025803}{\detokenize{10.1103/PhysRevC.97.025803}}}.

\bibitem[Walecka(1974)]{WALECKA1974491}
Walecka, J.
\newblock A theory of highly condensed matter.
\newblock {\em Annals of Physics} {\bf 1974}, {\em 83},~491 -- 529.
\newblock
  doi:{\changeurlcolor{black}\href{https://doi.org/https://doi.org/10.1016/0003-4916(74)90208-5}{\detokenize{https://doi.org/10.1016/0003-4916(74)90208-5}}}.

\bibitem[Johnson and Teller(1955)]{PhysRev.98.783}
Johnson, M.H.; Teller, E.
\newblock Classical Field Theory of Nuclear Forces.
\newblock {\em Phys. Rev.} {\bf 1955}, {\em 98},~783--787.
\newblock
  doi:{\changeurlcolor{black}\href{https://doi.org/10.1103/PhysRev.98.783}{\detokenize{10.1103/PhysRev.98.783}}}.

\bibitem[Glendenning(1997)]{norman1997compact}
Glendenning, N.K.
\newblock {\em Compact Stars: Nuclear Physics, Particle Physics, and General
  Relativity}; Astronomy and astrophysics library, Springer,  1997.

\bibitem[Schmitt(2010)]{Schmitt:2010}
Schmitt, A.
\newblock {Dense matter in compact stars: A pedagogical introduction}.
\newblock {\em Lect. Notes Phys.} {\bf 2010}, {\em 811},~1--111,
  \href{http://xxx.lanl.gov/abs/1001.3294}{{\normalfont
  [arXiv:astro-ph.SR/1001.3294]}}.
\newblock
  doi:{\changeurlcolor{black}\href{https://doi.org/10.1007/978-3-642-12866-0}{\detokenize{10.1007/978-3-642-12866-0}}}.

\bibitem[Meng(2016)]{meng2016relativistic}
Meng, J.
\newblock {\em Relativistic Density Functional for Nuclear Structure};
  International Review of Nuclear Physics, World Scientific Publishing Company,
   2016.

\bibitem[Prakash \em{et~al.}(1995)Prakash, Cooke, and
  Lattimer]{PhysRevD.52.661}
Prakash, M.; Cooke, J.R.; Lattimer, J.M.
\newblock Quark-hadron phase transition in protoneutron stars.
\newblock {\em Phys. Rev. D} {\bf 1995}, {\em 52},~661--665.
\newblock
  doi:{\changeurlcolor{black}\href{https://doi.org/10.1103/PhysRevD.52.661}{\detokenize{10.1103/PhysRevD.52.661}}}.

\bibitem[Wiringa \em{et~al.}(1988)Wiringa, Fiks, and
  Fabrocini]{PhysRevC.38.1010}
Wiringa, R.B.; Fiks, V.; Fabrocini, A.
\newblock Equation of state for dense nucleon matter.
\newblock {\em Phys. Rev. C} {\bf 1988}, {\em 38},~1010--1037.
\newblock
  doi:{\changeurlcolor{black}\href{https://doi.org/10.1103/PhysRevC.38.1010}{\detokenize{10.1103/PhysRevC.38.1010}}}.

\bibitem[Akmal \em{et~al.}(1998)Akmal, Pandharipande, and
  Ravenhall]{PhysRevC.58.1804}
Akmal, A.; Pandharipande, V.R.; Ravenhall, D.G.
\newblock Equation of state of nucleon matter and neutron star structure.
\newblock {\em Phys. Rev. C} {\bf 1998}, {\em 58},~1804--1828.
\newblock
  doi:{\changeurlcolor{black}\href{https://doi.org/10.1103/PhysRevC.58.1804}{\detokenize{10.1103/PhysRevC.58.1804}}}.

\bibitem[Özel and Freire(2016)]{Ozel:2016oaf}
Özel, F.; Freire, P.
\newblock {Masses, Radii, and the Equation of State of Neutron Stars}.
\newblock {\em Ann. Rev. Astron. Astrophys.} {\bf 2016}, {\em 54},~401--440,
  \href{http://xxx.lanl.gov/abs/1603.02698}{{\normalfont
  [arXiv:astro-ph.HE/1603.02698]}}.
\newblock
  doi:{\changeurlcolor{black}\href{https://doi.org/10.1146/annurev-astro-081915-023322}{\detokenize{10.1146/annurev-astro-081915-023322}}}.

\end{thebibliography}


\end{document}